\documentclass[lettersize,journal]{IEEEtran}
\usepackage{amsmath,amsfonts}
\usepackage{algorithmic}
\usepackage{algorithm}
\usepackage{array}
\usepackage[caption=false,font=footnotesize,labelfont=rm,textfont=rm]{subfig}
\usepackage{textcomp}
\usepackage{stfloats}
\usepackage{url}
\usepackage{verbatim}
\usepackage{graphicx}
\usepackage{cite}
\usepackage{amsthm,amssymb,bm}
\usepackage{subfloat}
\usepackage{float}
\usepackage{booktabs}
\usepackage{microtype}
\usepackage{lipsum}
\usepackage{cuted}
\usepackage{lineno}
\usepackage{xcolor}
\usepackage{cases}
\newcounter{MYtempeqncnt}

\newtheorem{lemma}{Lemma}

\begin{document}
\title{Integrated Communication and Positioning Design in RIS-empowered OFDM Systems: \\A Correlation Dispersion Scheme }

\author{Xichao~Sang,
        Lin~Gui,~\IEEEmembership{Member,~IEEE},
        Kai~Ying,~\IEEEmembership{Senior Member,~IEEE},
        Xiaqing~Diao,\\
        and Derrick Wing Kwan Ng,~\IEEEmembership{Fellow,~IEEE}
\thanks{X. Sang, L. Gui, K. Ying, X. Diao are with the Department of Electronic Engineering, 
Shanghai Jiao Tong University, Shanghai 200240, China 
(email: sang\_xc@sjtu.edu.cn; guilin@sjtu.edu.cn; yingkai0301@sjtu.edu.cn; 
xqdiao99@sjtu.edu.cn).}
\thanks{D. W. K. Ng is with the School of Electrical 
Engineering and Telecommunications, University 
of New South Wales, Sydney, NSW 2052, Australia 
(e-mail: w.k.ng@unsw.edu.au).}
\thanks{L. Gui is the corresponding author.}
}

\markboth{Journal of \LaTeX\ Class Files,~Vol.~XX, No.~X, February~2024}%
{Shell \MakeLowercase{\textit{et al.}}: A Sample Article Using IEEEtran.cls for IEEE Journals}


\maketitle

\begin{abstract}
  This paper proposes a novel reconfigurable intelligent 
  surface (RIS)-aided integrated communication 
  and positioning design for orthogonal frequency 
  division multiplexing systems in indoor scenarios. 
  A non-geometric strategy is employed to realize 
  accurate positioning. Specifically, location-related 
  information is embedded into channel frequency responses 
  (CFR) and estimated through regular pilot subcarriers. 
  The coefficients of RIS are optimized to maximize the 
  norm of  the CFR vector differences among users,  
  exclusively considering physically adjacent users. 
  To enhance positioning accuracy,  
  we propose a two-stage framework that incorporates the 
  prior information about the user in physical space.  
  A unique feature, named ``correlation dispersion'',
  within this framework is leveraged to enhance performance 
  compared to  geometric-based methods. By transforming the 
  geometric prior information into the frequency domain 
  capitalizing on Gaussian kernel method,  we derive the 
  Cramer-Rao Lower Bound (CRLB) of the proposed framework. 
  A notable gain in CRLB is observed, highlighting the efficacy.
  Theoretical comparison with the CRLB of conventional 
  methods validates the correlation dispersion property. 
  Simulation results demonstrate a significant improvement 
  in positioning accuracy when meticulously combining prior  
  information with a non-geometric positioning method.  
  Furthermore, our results unveil that the incorporation of 
  rough positioning methods yields exceptionally 
  high positioning performance,  
  provided that the location information depicts 
  different aspects.
\end{abstract}


\section{Introduction}
\begin{figure}[htb]
  \centering
  \includegraphics[width = 0.7\linewidth]{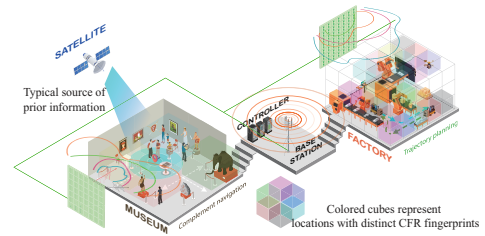}
  \caption{Potential application scenarios with different KPIs.}
  \label{scenarios}
\end{figure}

In recent years, there has been a surge in research 
directed towards various wireless technologies, 
including millimeter wave (mmWave), massive multiple-input multiple-output (MIMO), 
and reconfigurable intelligent surfaces (RISs), aimed at enhancing the coverage, 
throughput, and energy/spectral efficiency for numerous emerging applications \cite{10054381}. 
RISs, particularly, have garnered significant 
attention \cite{Basar, WUQ}, leveraging the  
programmability to facilitate information-bearing 
wireless signal propagation and giving rise to innovative concepts.
Noteworthy among these are the smart radio environment 
\cite{alexandropoulos_pervasive_2022} and the wireless 
environment as a service
\cite{strinati_reconfigurable_2021, strinati_wireless_2021}, 
both flourishing in the landscape 
of integrated sensing and communication (ISAC) \cite{10188491}.
It is widely anticipated  
that the introduction of RISs to wireless systems 
will enable dynamic electromagnetic wave control in the radio environment, 
facilitating customized services based on heterogeneous key 
performance indicators (KPIs)~\cite{strinati_reconfigurable_2021}. 
This development aligns perfectly with the future vision 
of location/environment-aware applications~\cite{9625159}, 
e.g., seamless navigation transition, accurate indoor trajectory planning,
as illustrated in Fig.~\ref{scenarios}. 
These applications are envisioned for 
beyond-fifth-generation (B5G) and sixth-generation~(6G) 
wireless networks~\cite{strinati20216g}.

As one of the most crucial elements of interconnecting and sensing,
precise positioning is a key factor and an 
enabling technology for various use cases
in the ISAC systems.
In relation to this, the exploration of 
integrated communication and positioning
(ICAP)
has received extensive attention
in both the academic and industrial sectors \cite{koirala_localization_2017,koirala_localization_2019,
xu2017accuracy,Resonant_Beam,williams2018impact,3GPP_R16}.
The fundamental principle of ICAP is grounded in 
the shared characteristic that both wireless communication 
and wireless positioning extract information from the traversing
of electromagnetic beams and other signals \cite{10086594}.
As such,
aligned with the promising design of ICAP for future wireless networks, 
substantial progress has been made in early research. 
Notably, the authors in~\cite{koirala_localization_2017} 
proposed a comprehensive resource allocation strategy
for ICAP in mmWave frequency bands. This strategy involves 
estimating the angle-of-arrival (AoA) for positioning and 
developing a protocol for the allocation of 
time and frequency domain resources.
To further enhance performance of the frequency selection strategy, 
the strategy in \cite{koirala_localization_2019} improved the 
approach by considering the quality of service (QoS) 
of heterogeneous services, aiming at maximizing the 
overall transmission rate.
Moreover, the authors in
\cite{xu2017accuracy} utilized visible light 
communication (VLC) systems and exploited the detection 
of time-difference-of-arrival (TDoA) for positioning. 
It proved that orthogonal frequency division multiplexing (OFDM)-based 
system was compatible with 
the triangulation positioning.
More recently, a monocular resonant beam-based design 
was proposed in~\cite{Resonant_Beam}. This design 
aims to improve the directivity of antenna beams,
thereby enhancing positioning 
beyond VLC technology.

In the aforementioned integration explorations, 
the crux for achieving accurate positioning 
is the avoidance of complex multipath circumstances, 
as the indistinguishable link may attenuate accuracy. 
The authors exclusively assumed that the propagation process 
is dominated by line of sight (LoS), 
simplifying the abstraction of location-related 
information \cite{koirala_localization_2017,koirala_localization_2019,
xu2017accuracy,Resonant_Beam}. However, such an assumption is unrealistic, 
especially when dealing with practical wireless communication, 
as non-LoS links often offer additional gain to 
improve channel capacity.
Among existing technologies, RIS, featured for reconfiguring
the propagation process, 
introduces an additional degree of freedom (DoF)
to harmonize these aspects of communication and positioning through the modulation 
of the propagation channel.

Leveraging the extra DoF provided by RIS,
the authors in \cite{9215972} 
circumvented blockages through the reflected 
link and treated RIS as an anchor
to realize positioning through AoA estimate.
By creating a strong and consistent multipath, 
RISs can support localization in even very harsh indoor environments.
Inspired by the RIS's ability to alleviate the impairment caused by multipath, 
the authors in \cite{9124848} proposed a strategy to align all non-LoS links with
the LoS to enhance the positioning accuracy.
Moreover, the authors 
in \cite{nguyen_wireless_2021} combined
the traditional received signal strength (RSS)-based 
localization approach with RIS by designing a machine 
learning-based approach to select optimal RIS phases 
and artificially create fingerprint libraries to 
achieve localization. 
Also, the authors in \cite{zhang_rss_2021,zhang_metaradar_2022} used a 
probabilistic analytical-based approach to improve 
the localization accuracy by optimizing the localization error probability and dividing 
the localization process into multiple time slots.

While existing RIS-based designs showcase 
impressive positioning performance, typical RIS-aided methods 
still fall short of fully resolving the 
drawback imposed by multipath \cite{9215972,9124848}. 
The fundamental challenge stems from the 
fact that common positioning technologies, 
AoA, TDoA, time-of-arrival (ToA), and RSS, 
are geometric-based. 
This implies that the exact location is not 
directly obtained. Thus, as long as all the
links, including LoS link and non-LoS links,
are not fully exploited, there is 
a loss of information in positioning \cite{kay1993fundamentals}.
However, non-LoS links always introduce extra interference 
to positioning due to the existence of 
spatial consistency \cite{8926349}.
Here, spatial consistency or spatial correlation 
refers to the similarity in propagation
that closely located users experience similar 
channel propagation effects.
Notably, the fingerprint-based methods proposed in 
\cite{nguyen_wireless_2021} and \cite{zhang_metaradar_2022} 
provide an early exploration by considering 
all the channel information to construct a radio map.
By collecting snapshots of every configuration 
of RIS, each location possesses a unique fingerprint for 
localization. The obtained information is not geometric-based, 
thus achieving even more precise localization benefiting from a
different spatial correlation.

However, positioning through a series of modulations of RIS 
in \cite{nguyen_wireless_2021} and \cite{zhang_metaradar_2022}
represents a form 
of diversity in the time domain. Apart from being time-consuming, 
this approach may jeopardize the communication 
requirements of the inherent system.
Furthermore, all the aforementioned 
RIS-aided positioning designs assume perfect channel state information (CSI) 
at the base station (BS), which is impracticable due to 
the limitations of the RF chain equipped at the RIS.
A robust design approach assumption is often necessary, 
considering the time-varying nature 
of the location and the inherent 
inaccuracies in the channel estimate from RIS.

Therefore, in this paper, our objective is to propose a holistic
RIS-aided ICAP
design
that seamlessly integrates communication and positioning functionalities
in the presence of channel uncertainty. 
Built upon the concept that RIS can act as a spatial filter to 
mitigate the frequency selective fading
and multipath effect\cite{9414612, arslan2022over}, 
we further investigate capability of RIS to
reshape spatial correlation dispersion.
We propose that RIS can be exploited to modulate 
the frequency characteristics of the radio environment 
by altering the spatial distribution of propagating waves. 
Based on the location-related information above, we design a 
two-stage positioning method with a single-time
configuration of RIS instead of extensive searches across multiple time slots 
\cite{nguyen_wireless_2021,zhang_rss_2021,zhang_metaradar_2022}.
To be specific, the channel frequency response (CFR) of pilots of each place of interest (POI) 
that forms a vector can be treated as 
the corresponding virtual coordinate in the virtual positioning domain. That
is to say, the exact location of the POI in the physical space is mapped into a virtual space 
through a linear transformation introduced by RIS.
As such, the objective of minimizing the similarity of different CFRs is 
transformed into optimally designing the pattern of distribution of the virtual coordinates.
We prove theoretically that the positioning accuracy
is significantly improved through the combination
of two positioning spaces that share distinct spatial consistency.
The main contributions of this
work are summarized as follows:
\begin{itemize}
  \item We introduce a novel integrated communication and localization scheme in the
  context of a RIS-assisted OFDM communication system.
  RIS is leveraged to introduce enhanced location-related information.
  Notably, we enhance both communication and positioning function 
  capability without additional 
  resources or modifications to the transceiver system.
  
  \item A two-stage positioning framework that innovatively incorporates the prior information
  is proposed to enhance the positioning performance.
  A novel characteristic, named ``correlation dispersion'',
  has been identified, indicating that the spatial consistency 
  can be redesigned to construct a positioning-favor space with the help of RIS. 
  This property enables extremely high 
  positioning accuracy, particularly in low signal-to-noise ratio (SNR) conditions.

  \item Theoretical interpretation of 
  the proposed correlation dispersion property is presented.
  The prior information is transformed into the frequency domain utilizing the 
  Gaussian kernel method
  and the corresponding
  Cramér-Rao Lower Bound (CRLB) of the proposed positioning framework is derived.
  Compared to conventional geometric-based methods \cite{zhang_rss_2021}, 
  our scheme exhibits a 
  significant reduction in CRLB.

  \item Simulation results validate the effectiveness of the proposed 
  joint communication and localization scheme.
  The results demonstrate a great improvement in low-SNR conditions
  compared to conventional methods.
\end{itemize}

\subsection{Organization \& Notations}
The rest of this paper is organized as follows. 
In Section II, we present the system model and formulate the problem 
of constructing the positioning space via a RIS.
Section III introduces a general algorithm designed to address the non-convex 
optimization problem. Also,
a novel property, termed ``correlation dispersion'', is firstly proposed.
In Section IV, we derive the CRLB for the proposed 
positioning method and provide a theoretical interpretation of 
correlation dispersion. Simulation results are presented in Section V 
and Section VI concludes the paper.

The notations used in this paper are listed as follows.
Upper and lower case boldface letters denote
matrices and column vectors, respectively.
$\mathbb{E}[\cdot]$ stands for statistical expectation. 
$\mathbb{C}^{M\times N}$ denotes the $M\times N$ complex-valued matrix.
$\mathbf{I}_M$ denotes the $M \times M$ 
identity matrix. For any general matrix $\mathbf{A}, A_{i, j}$ is 
the $i$-th row and $j$-th column element. $\mathbf{A}^*, 
\mathbf{A}^{\mathrm{T}}$, and $\mathbf{A}^{\mathrm{H}}$ denote the 
conjugate, the transpose, and the conjugate transpose of $\mathbf{A}$, 
respectively. 
For any vector, $\bf{x}$ (all vectors in this paper 
are column vectors), $X_{i-1}$ is the $i$-th element.
$\operatorname{diag}(\mathbf{x})$ returns a
diagonal matrix whose diagonal elements are included in $\bf{x}$,
$\operatorname{blkdiag}(\cdot)$ returns a block diagonal matrix 
created by aligning the input matrices or vectors.
Reversely, $\operatorname{Diag}(\mathbf{A})$ yields a vector 
containing the diagonal elements of matrix $\mathbf{A}$.
$\mathbf{A} \succeq \mathbf{0}$ implies that $\mathbf{A}$
is positive semidefinite.
$\operatorname{Tr}(\cdot)$ returns the trace of the matrix. 
$\operatorname{rank}(\cdot)$ represents the rank of the input matrix.
$\circ$ denotes the element-wise
produce.
$\|\cdot\|_2$ and $\| \cdot \|_{\mathrm{F}}$ denote 
the $\mathbb{\ell}_2$-norm and the Frobenius norm, respectively. $|x|$ 
denotes the modulus of a complex number $x$.

\section{System Model and Problem Formulation}
This paper investigates an indoor RIS-aided MISO-OFDM communication system, 
whose channel model follows the approach  
in \cite{lin_adaptive_2020}.
The system structure is illustrated in Fig. \ref{scenarios}, 
where the colored areas most likely to 
have users within the cell are designated as places of interest (POIs). 
These areas are divided into $I$ distinct blocks.
The centers of these blocks serve as the sampling locations, 
referred to as $\mathcal{I} =\{ 1,2,\cdots, I\} $.
Typically, the features of received wireless signals serve 
as the basis of location distinction and 
determination for wireless indoor localization~\cite{yang_rssi_2013}.
This paper adopts the concept of CFR fingerprints of users for positioning.
The core idea of the fingerprint-based method 
involves matching a CFR vector 
in a database to an estimated CFR vector.
Unlike traditional fingerprint measurements,
which suffer from severe attenuation due to the multipath 
effects of signal propagation \cite{tse2005fundamentals}, 
the proposed RIS-aided 
positioning method focus on reshaping
the multipath distribution, 
specifically the distribution of CFR. 
Our objective is to construct a communication-compatible 
fingerprint database with high resolution.


\subsection{System Model}
We consider a RIS-assisted transmission in the downlink, where a 
broadcast channel is employed. 
The system utilizes a RIS composed of $M$ units of reflective elements, 
indexed by $\mathcal{M} =\{ 0,1,\cdots, M-1\} $, to modulate the 
transmission from a BS equipped with $N_\mathrm{T}$ antennas to 
$K$ single-antenna users, denoted as $\mathcal{K} =\{0,1,\cdots, K-1\} $, requiring the same 
piece of information.
The total bandwidth for transmission
allocated to the users is uniformly divided into $N$ orthogonal subcarriers, denoted by the 
index set $\mathcal{N} \triangleq \{0,1,\cdots , {N-1} \}$.
Specifically, we define $\mathbf{h}_{k,n}\in \mathbb{C}^{N_{\mathrm{T}}\times 1}$ 
and $\mathbf{G}_{k,n}
\in \mathbb{C}^{N_{\mathrm{T}}\times M}$ as the 
channel matrices representing the direct and the cascaded links, respectively, 
for the $k$-th user on the $n$-th subcarrier, $n\in \mathcal{N}$.
The reflection performed  on the $m$-th RIS element
is to multiply the incident signals with $\theta_m$, $m\in\mathcal{M}$,
and then
forward the composite signals, where
$\theta_m=e^{-\jmath \phi_m}$ denotes the  
reflection coefficient of 
the $m$-th element with phase shift $\phi_m\in (0,2\pi]$.

At the BS side, we assume that the transmitted 
information across the $N$ subcarriers 
is given by $\mathbf{s}=[s_{0}, s_{1}, 
\cdots, s_{N-1}]^\mathrm{T} \in \mathbb{C}^{N\times 1}$,
satisfying $\mathbb{E}[\mathbf{s}\mathbf{s}^\mathrm{H}]=\mathbf{I}_N$.
Prior to transmission, the signal over $N_\mathrm{T}$ antennas on the $n$-th subcarrier
is firstly digitally precoded by a beamforming vector as
\begin{align}
  \mathbf{x}_n = \mathbf{f}_ns_n,
\end{align}
where $\mathbf{f}_n\in \mathbb{C}^{N_\mathrm{T}\times 1}$ 
is the beamforming vector on the $n$-th subcarrier.
In this paper, our focus is to 
investigate the RIS's potential for positioning. 
The beamforming of the antenna array will be 
determined using traditional methods such as maximum ratio transmission (MRT), 
minimum mean square error (MMSE) or other algorithms\footnote[1]
{This paper aims to explore the inherent positioning functionality facilitated 
by RIS and to assess the achievable positioning performance through RIS modulation.
While dynamic beamforming techniques can introduce unpredictable channel 
fluctuations that may impact positioning, we employ a fixed beamforming 
scheme to exclusively investigate 
RIS-based positioning, ensuring a focused analysis. In the future work, we will
extend to optimize both the phase shifts of RIS and 
precoders at the BS.} 
\cite{5756489, 7397861}.
Subsequently, each OFDM symbol denoted by $\mathbf{X} =\operatorname{blkdiag}
([\mathbf{x}_0,\mathbf{x}_1,
\cdots,\mathbf{x}_{N-1}])\in \mathbb{C}^{NN_\mathrm{T}\times N}$ 
is first transformed into the time domain
via an $N$-point inverse discrete Fourier transform (IDFT),
and then appended with a cyclic prefix (CP)
to avoid inter-symbol interference.
At the user side, the received signal undergoes CP removal and is 
subjected to an $N$-point discrete Fourier transform (DFT) \cite{feng_joint_2021}. 
The received baseband signal at the $k$-th user, $k\in\mathcal{K}$,
in the frequency domain over the $N$ subcarriers is then 
given by
\begin{align}
  \label{eq:revdSig}
  \mathbf{y}_k=\mathbf{X}^\mathrm{H}
  \left(\tilde{\mathbf{G}}_{k}\boldsymbol{\theta}+
  \tilde{\mathbf{h}}_{k}\right)+\mathbf{n}_k.
\end{align}
We define $\tilde{\mathbf{G}}_{k}\triangleq
[\mathbf{G}_{k,0}^\mathrm{H}, \mathbf{G}_{k,1}^\mathrm{H}, 
\cdots, \mathbf{G}_{k,N-1}^\mathrm{H}]^\mathrm{H}\in \mathbb{C}^{NN_\mathrm{T}\times M}$ and 
$\tilde{\mathbf{h}}_{k} \triangleq [\mathbf{h}_{k,0}^\mathrm{H}, \mathbf{h}_{k,1}^\mathrm{H}
\cdots, \mathbf{h}_{k,N-1}^\mathrm{H}]^\mathrm{H}\in \mathbb{C}^{NN_\mathrm{T}\times 1}$  
as the concatenated channel matrix/vector for the cascaded links and direct link, respectively. 
$\boldsymbol{\theta} = [\theta_{0},\theta_{1},\cdots,\theta_{M-1}]^\mathrm{T}
\in \mathbb{C}^{M\times 1}$ is 
the phase shifts vector of the RIS.
We consider the presence of identical additive white Gaussian noise (AWGN) 
across different users and subcarriers
as $\mathbf{n}_k\sim \mathcal{CN}(\boldsymbol{0},\sigma^2_0\mathbf{I}_N)$,
where $\sigma_0^2$ is power of noise on each subcarrier.
In this way, the end-to-end designed channel frequency response (CFR) of the $k$-th user is represented as
\begin{equation}
  \begin{aligned}
    \label{eq:CFR}
    \mathbf{r}_k\triangleq & [R_{k,0},R_{k,1},\cdots, R_{k,N-1}]^\mathrm{T}
    \\=&\mathbf{F}^\mathrm{H}\tilde{\mathbf{G}}_{k}\boldsymbol{\theta}+
    \mathbf{F}^\mathrm{H}\tilde{\mathbf{h}}_{k}
    =\mathbf{F}^\mathrm{H}{\mathbf{G}}'_k\boldsymbol{\theta}', k\in\mathcal{K},
  \end{aligned}
\end{equation}
where $\mathbf{F} \triangleq \operatorname{blkdiag}
([\mathbf{f}_0,\mathbf{f}_1,
\cdots,\mathbf{f}_{N-1}])\in \mathbb{C}^{NN_\mathrm{T}\times N}$ is the beamforming
matrix over all subcarriers.
Augmented matrix ${\mathbf{G}}'_k \triangleq \left[
\tilde{\mathbf{G}}_{k}, \tilde{\mathbf{h}}_{k}
\right]\in\mathbb{C}^{NN_\mathrm{T}\times (M+1)}$ and augmented 
vector ${\boldsymbol{\theta}'}\triangleq \left[
{\boldsymbol{\theta}}^\mathrm{T}, t\right]^\mathrm{T}
\in\mathbb{C}^{(M+1)\times 1}$
are introduced for ease of analysis in the sequel.
Here, $t$ is an auxiliary variable.
Once the optimal solution is found, ${\boldsymbol{\theta}}$ can
be recovered by ${\boldsymbol{\theta}} = 
\exp\left({\boldsymbol{\theta}'_\mathcal{M}}/{t}\right)$.

\begin{figure}[!tb]
  \centering
  \includegraphics[width = 0.85\linewidth]{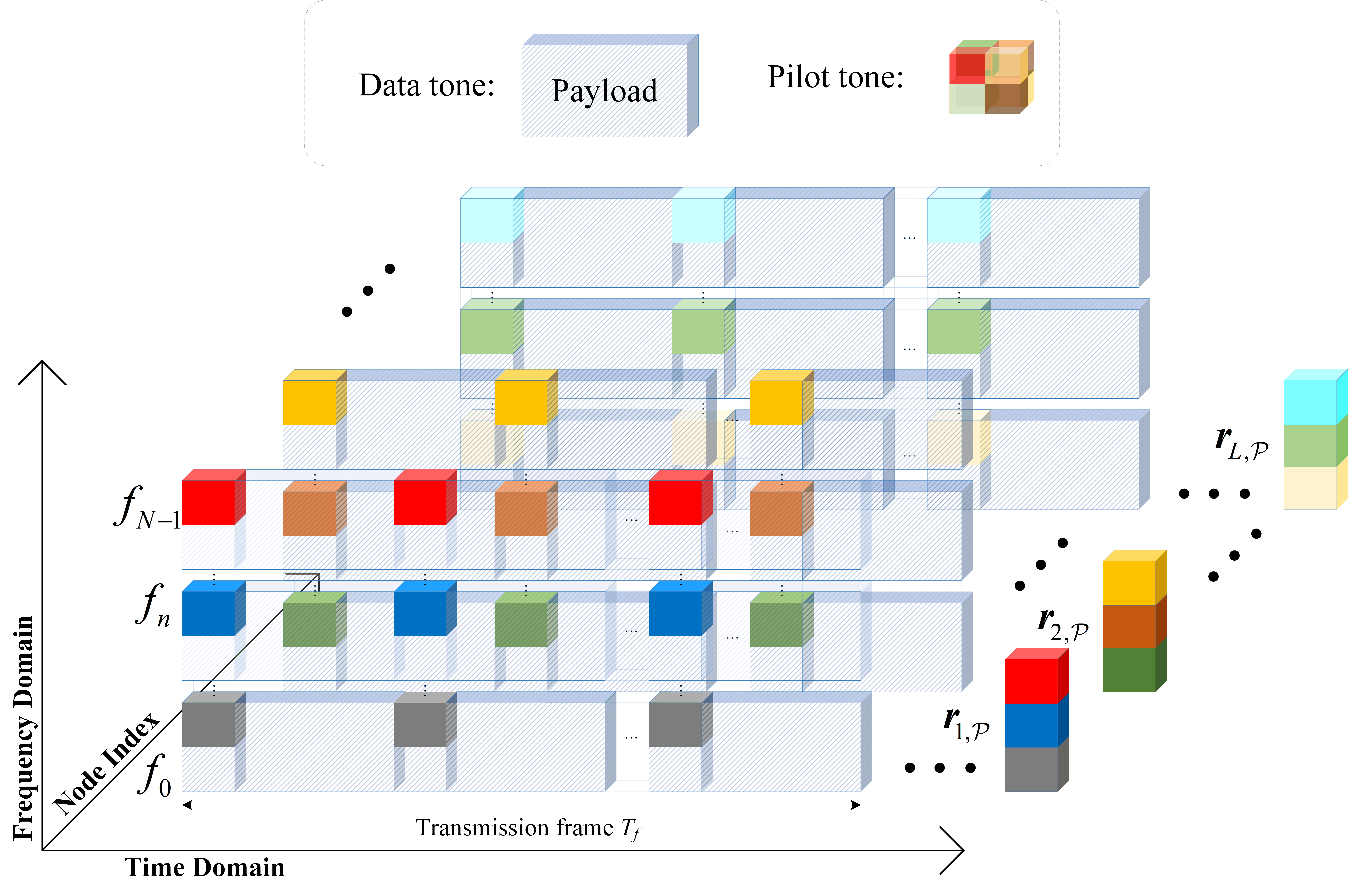}
  \caption{Resource blocks with colored pilots.}
  \label{RBs}
\end{figure}
To embed the positioning function through RIS, 
we leverage the existing training pilot sequence \cite{kay1993fundamentals}.
As illustrated in Fig. \ref{RBs},
we assume that a total of $N_p$ pilots are uniformly inserted
within each training OFDM symbol.
These pilots are indexed by 
$\mathcal{P} = \{0, \Delta_p, \cdots, (N_p-1)\Delta_p\}$,
where $\Delta_p = \left \lfloor N/N_p \right \rfloor $ 
is the frequency spacing of adjacent pilots. 
By exploiting the public unit-power pilot sequence $\mathbf{s}\in\mathbb{C}^{N_p\times 1}$, 
it becomes possible to estimate the CFR
of the $k$-th user on the pilot subcarriers $\mathcal{P}$.
The corresponding subvector $\hat{\mathbf{r}}_{k,\mathcal{P}}\in\mathbb{C}^{N_p\times 1}$ 
that includes the CFR on pilot subcarriers is expressed as
\begin{align}
  \label{eq:cfrvec}
  \hat{\mathbf{r}}_{k,\mathcal{P}} = \mathbf{S}^{-1}\mathbf{y}_{k,\mathcal{P}}
   = \mathbf{r}_{k,\mathcal{P}} + \mathbf{S}^{-1}\mathbf{n}_{k,\mathcal{P}},
\end{align}
where $\mathbf{S}=\operatorname{diag}({\mathbf{s}})$ 
is obtained by the diagonalization of the pilot sequence, 
and $\mathbf{y}_{k,\mathcal{P}}\in\mathbb{C}^{N_p\times 1}$ is the extracted received signal on 
pilot subcarriers.
For ease of analysis, we assume that the updated noise follows 
isotropic distribution with unchanged power.
In the subsequent discussion, 
with a slight abuse of notation,
the subscript $n$ is utilized 
consistently to denote the pilot index within~$\mathcal{P}$.
Consequently, CFR and the corresponding channel matrices are 
consistently referred to as the CFR on the pilot carriers,
unless specified otherwise.

\subsection{Two-Stage Positioning Framework}
\begin{figure}[tb]
  \centering
  \includegraphics[width=0.71\linewidth]{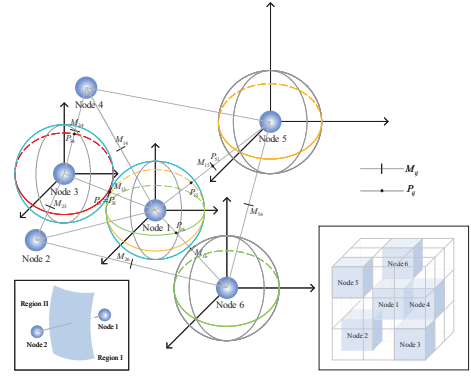}
  \caption{Mapping from the physical domain to the topological domain.}
  \label{fig3}
\end{figure}
In this paper, we utilize the commonly adopted scaled-$\ell_2$-norm-with-log-probability
for localization, which is derived 
from maximizing a posterior probability \cite{torres2015comprehensive}, expressed as
\begin{equation}
  \begin{aligned}
    \label{eq:ell2P}
    i_0=&\arg \max_{i} P(\hat{\mathbf{r}}_{\mathcal{P}}|{\mathbf{r}}_{i,\mathcal{P}})\\
    \Leftrightarrow 
    i_0=&\arg \max_{i}
    \left[-\left\|\hat{\mathbf{r}}_{\mathcal{P}}-\mathbf{r}_{i,\mathcal{P}}\right\|_2^2/\sigma_0^2
     +2\log P({\mathbf{r}}_{i,\mathcal{P}})\right],
  \end{aligned}
\end{equation}
where $\hat{\mathbf{r}}_{\mathcal{P}}$ is the estimated CFR 
vector through \eqref{eq:cfrvec}.
$P({\mathbf{r}}_{i,\mathcal{P}}|\hat{\mathbf{r}}_{\mathcal{P}})$
and $P({\mathbf{r}}_{i,\mathcal{P}})$ are
the posterior probability given the 
estimated CFR vector and the prior 
probability of the user at the $i$-th block, respectively.
The prior information 
can be readily collected from previous estimates or
GNSS~\cite{9330772}, typically providing a coarse estimate.
However, as both prior information and fingerprint 
data originate homogeneously from the geometric space, 
the prior information contributes 
limited distinctive location details.
This homogeneity implies that inaccuracies in prior 
information align with those in fingerprint positioning, 
particularly in proximity to the given location.

To break through the limitation, 
we propose a two-stage framework.
In the initial \textbf{prediction stage},
the prior information is employed to predict and narrow down the potential
range of user locations over a certain confidence probability.
Instead of pinpointing the exact location, we consider all locations within 
this confidence range without any bias. 
In the subsequent \textbf{correction stage}, 
we utilize the embedded CFR fingerprint information to determine the precise 
location among those within the confidence range established in the first stage.
In particular, the distribution within the database of all 
fingerprints undergoes reorganization through RIS.
This aims to exclude all potential locations that are physically 
close to the exact location so that the fingerprint information 
exhibits diverse spatial consistency compared to the prior information.

As exemplified in Fig.~\ref{fig3},
each user's location are mapped into two spaces, i.e., the physical space and the 
virtual positioning space. Specifically
the virtual positioning domain is constructed by treating 
the CFR subvector as the corresponding coordinate.
For instance, the coordinate of a user at the $i$-th block
is denoted as $(\mathbf{F}^\mathrm{H}{\mathbf{G}}'_i\boldsymbol{\theta}')_\mathcal{P}$,
where the operation $(\cdot)_\mathcal{P}$ yields the subvector or submatrix 
comprising of the elements or columns indexed by specified subcarriers.
Note that these coordinates do 
not directly convey the geometric properties of the user's location. 
Instead, they serve as a mapping of the sample index $i \in \mathcal{I}$ to the
positioning space. 
The dimension of the virtual positioning space is consistent with the
dimension of CFR.

In Section IV, we will conduct an examination on the 
CRLB pertaining to the positioning scheme that integrates the prior information. 
Furthermore, we will demonstrate
that the positioning performance of the proposed two-stage framework 
can achieve the corresponding CRLB.

\subsection{Channel Uncertainty}
Due to the limitations of RF chain equipped at RIS and signal processing 
capability in exploiting previous transmission 
slots for estimation, obtaining accurate CFR 
is challenging \cite{9149292}. In this paper, we assume that the CFR estimated 
at the user side experiences some stochastic but bounded errors.
By adopting a deterministic model, 
it becomes possible to characterize the channel model 
with inherent uncertainties as \cite{Derrick}:
\begin{align}
  \label{eq:ICSI}
  {\mathbf{G}}'_{k} &= \hat{\mathbf{G}}'_{k}+ \Delta{\mathbf{G}}'_{k},
  k\in\mathcal{K},\\
  \boldsymbol{\Omega}_{k} &\triangleq \left\{\Delta{\mathbf{G}}'_{k}:
  \left\|\Delta{\mathbf{G}}'_{k}\right\|^2_\mathrm{F} 
  =\sum_{n\in\mathcal{N}} \|\Delta{\mathbf{G}}_{k,n}\|_\mathrm{F} ^2
  \leq \sum_{n\in\mathcal{N}}\epsilon_{k,n}^2
  = \epsilon_{k}^2 \right\},
\end{align}
where $\hat{\mathbf{G}}'_{k}$ denotes the estimate of the actual CSI, 
while $\Delta{\mathbf{G}}'_{k}$ and
$\Delta{\mathbf{G}}_{k,n}$ represent random estimation errors of 
the channel matrix in \eqref{eq:CFR} and cascaded channel of the $k$-th user
on the $n$-th subcarrier, respectively.
Non-negative factor $\epsilon_k$ and $\epsilon_{k,n}$ denote the 
bound of estimation uncertainty of $\Delta{\mathbf{G}}'_{k}$ and
$\Delta{\mathbf{G}}_{k,n}$.
Note that the LoS is perfectly estimated and 
the channel uncertainty assumption primarily pertains to
the estimation error from the cascaded link. 
The set $\boldsymbol{\Omega}_k$ is defined as a continuous set encompassing 
all potential CSI estimation errors, with their norms bounded by 
$\epsilon_k$. We assume that the BS can acquire the knowledge of 
$\hat{\mathbf{G}}'_k$ and ${\boldsymbol{\Omega}_k}$, 
which refers to the estimates of the CSI and the corresponding 
uncertainty for the $k$-th user.

\subsection{Problem Formulation}
According to the positioning criterion in \eqref{eq:ell2P}, 
a greater separation between any two physically close locations 
consistently results in better positioning accuracy. 
In the context of our proposed two-stage framework, 
a straightforward strategy involves expanding among 
locations within the potential area over the virtual
positioning space. This ensures that each individual 
location can be accurately identified for a given energy of additive noise.
As such, the subproblem of maximizing the worst-case 
positioning accuracy can be expressed as
the mostly used $\ell_2$-norm of the 
coordinate differences:
\begin{IEEEeqnarray}{lcll}
  \mathcal{P}_0\text{:}~ &  \underset{d, \boldsymbol{\theta}'}{\max }         \min_{\Delta\mathbf{G}', i, j}  \  & 
                            \left\|{{\mathbf{r}}_{i,\mathcal{P}} 
                            -{\mathbf{r}}_{j,\mathcal{P}}}\right\|^2_2,
\end{IEEEeqnarray}
where the minimization operation over the channel uncertainty 
aims to ensure an optimum for worst-case channel realizations.


Based on the primary focus of this paper to 
establish a RIS-aided integrated communication and positioning framework,
our objective is to determine a set of optimal phase 
coefficients of RIS that
maximizes positioning accuracy constrained by communication quality requirements.
In the domain of RIS-aided communication, 
the optimization of the achievable data rate
is a prevalent focus
\cite{yang_qos-driven_2020}.
Based on the transmission system in this paper, the achievable data rate of 
the $k$-th user over $N$ subcarriers can be expressed as:
\begin{equation}
  \label{eq:datarate}
  C_k(\boldsymbol{\theta})=
  \sum_{n\in\mathcal{N}} \log_2\left(1+
  \frac{{W}_{k,n}(\boldsymbol{\theta}
  )}{\sigma_n^2}\right),
\end{equation}
where ${W}_{k,n}(\boldsymbol{\theta})\triangleq
\left|\mathbf{f}_n^{\mathrm{H}}(\mathbf{G}_{k,n}\boldsymbol{\theta}+
\mathbf{h}_{k,n}) \right|^2$ 
is the channel gain of the $k$-th user on the $n$-th subcarrier.
The data rate, $C_k(\boldsymbol{\theta})$,
is measured in bits per second per Hertz (bps/Hz).
To uphold the desired communication requirement,
we impose a lower bound constraint for the 
achievable data rate~\cite{6810493} as 
$C_{k}(\boldsymbol{{\theta}}) \geq R_k$,
where $R_k$ is the minimum requirement.

To facilitate the optimization between communication and positioning, 
we introduce an auxiliary optimization variable, 
denoted as $d$, representing a lower bound for the mentioned $\ell_2$-norm in $\mathcal{P}_0$.
The optimization problem can then be formulated in the epigraph form as
\begin{IEEEeqnarray}{lcll}
  \mathcal{P}_1\text{:}~ &  \underset{d, \boldsymbol{\theta}'}{\max }   ~                                               & d                                                               & 
  \label{eq:origin_P}  \\
  &\text{s.t.}                                                                       & \min_{\Delta\mathbf{G}'} \left\|{{\mathbf{r}}_{i,\mathcal{P}} 
  -{\mathbf{r}}_{j,\mathcal{P}}}\right\|^2_2\geq d, \forall i,\in \mathcal{I},j\in \mathcal{U}_i,
  \IEEEyessubnumber\label{eq:huge_constraint1}\\
  &                                                                                  & \min_{\Delta\mathbf{G}'}C_{k}(\boldsymbol{{\theta}}) \geq R_k,                                              \forall k \in \mathcal{K},
                      \IEEEyessubnumber\label{eq:SNRconstraint}\\
                      &                                                                                  & \label{unit constraint} |{{\theta}}_{m}|= 1,                      \forall m \in \mathcal{M},
                      \IEEEyessubnumber \label{eq:unit_constraint}
\end{IEEEeqnarray}
where the set $\mathcal{U}_i$ includes those locations close to location $i$ within 
a maximum physical distance of $\delta$, exclude $i$ itself. Here, $\delta$ is  
determined based on the confidence probability mentioned in the prediction stage 
of the proposed positioning framework.
Based on the assumption that the additive noise is isotropically 
Gaussian distributed, the following equation holds:
\begin{align}
  \delta\triangleq \sqrt{2}\operatorname{erfc}^{-1}\left( 
    \frac{1-\sqrt[3]{\alpha}}{\sigma}
  \right),
\end{align}
where $\operatorname{erfc}^{-1}(\cdot)$ is the inverse function of the complementary
error function.
Note that the problem $\left(\mathcal{P}_1\right)$ is positioning-centric, 
which aligns with the primary focus of this paper on assessing performance. 
The problem is challenging due to the infinite constraints 
arising from continuous channel uncertainty and the non-convex 
constraints \eqref{eq:SNRconstraint} and \eqref{eq:unit_constraint}. 
In the next sections, we will address these challenges by  
transforming the stochastic constraints 
into finite linear matrix inequalities (LMIs) 
and handling the non-convex constraint 
through approximation.

\section{Algorithm Design}

\subsection{Correlation Dispersion}
\label{secIIIB}
In formulating the optimization problem, our focus is exclusively on the 
neighborhood of each location for positioning design.
This selection aligns with the first phase of the proposed 
two-stage positioning framework, where we exclude 
the distant locations. 
For conventional geometric-based positioning methods \cite{9215972,9124848}, 
inaccuracies in estimating locations among nearby areas always 
incur significant errors in positioning results. 
In contrast, locations farther away from that are less likely to contribute to 
significant mislocalization issues. 
Inspired by this observation,
we propose the following proposition: the positioning accuracy will be significantly 
improved if there are different descriptions of location information that exhibit 
varied spatial consistency. 
We refer to the characteristic that reshapes the spatial 
consistency as ``correlation dispersion''.

In the first phase of our two-stage framework, 
we incorporate the geometric prior information to exclude
physically separated locations.
In the second phase, we aim to enhance the accuracy 
of localization in the vicinity of each point 
by prioritizing physically close locations
in the virtual positioning space.
Next, we qualitatively exemplify the correlation dispersion property
and then theoretically 
prove it in Section~IV.


\begin{figure}[htb]
  \centering
  \includegraphics[width=0.65\linewidth]{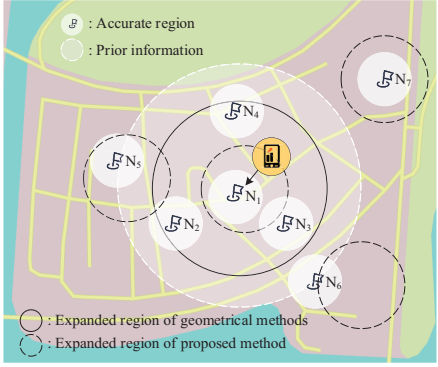}
  \caption{Robust localization when affected with noise.}
  \label{fig4}
\end{figure}
As depicted in Fig. \ref{fig4}, we aim to estimate the user's location within 
the range of $N_1$
under challenging situations.
The white colored circles represent the accurate positioning 
zones with a 95\% confidence probability for each block, 
while the circle with a white dashed border 
indicates the potential zone 
where the user is likely to appear based on the prior information.
Conventional techniques such as angular estimating methods may 
encounter severe errors \cite{3GPP_R16},
causing the potential positioning range (circles with black solid borders) 
to expands, leading to a degradation 
in positioning accuracy. 
Such methods can only infer that the user's 
location is within the black-border circle without specific information about 
regions of $N_2$, $N_3$, or $N_4$,
making the prior information redundant.

By contrast, our proposed scheme smartly utilizes
the prior information as an additional reference in an another domain.
Despite the potential positioning range
(circles with black-dashed borders) expands due to noise,
inaccuracy is dispersed into the entire virtual positioning space, 
as illustrated in Fig.~\ref{fig3}. 
With the coarse prior information, we identify the user's location by 
considering the intersection of the white-dashed circle and the black-dashed circles. 
Consequently, maintaining or improving the accuracy in challenging environment.

\subsection{S-procedure}
To handle the optimization problem,
we first utilize matrix operations to express the channel 
difference matrix in a concise form as ${\mathbf{D}}_{ij} \triangleq {\mathbf{D}}_{i} - 
{\mathbf{D}}_{j} = \mathbf{F}^{\mathrm{H}}(\mathbf{G}^\prime_i-\mathbf{G}^\prime_j)$.
$\hat{\mathbf{D}}_{ij}\triangleq 
\hat{\mathbf{D}}_{i}-\hat{\mathbf{D}}_{j}$ 
and $\Delta{\mathbf{D}}_{ij}\triangleq \Delta{\mathbf{D}}_{i}-\Delta{\mathbf{D}}_{j}$
are defined in a similar manner.
To characterize the uncertainty of channel difference matrices,
we introduce $\boldsymbol{\varOmega_{ij}}$ defined as 
$\boldsymbol{\varOmega_{ij}}\triangleq \{\Delta{\mathbf{D}}_{ij}:
\left\|\Delta{\mathbf{D}}_{ij} \right\|^2_\mathrm{F}  
\leq \hat{\epsilon}^2_{ij}\}$. Specifically, $\hat{\epsilon}_{ij}$ is obtained by
\begin{equation}
  \begin{aligned}
    \label{eq:nor_uncertainties}
    \left\|\Delta{\mathbf{D}}_{ij} \right\|^2_\mathrm{F} 
    &=\left\|{\mathbf{F}}^\mathrm{H}\Delta{{\mathbf{{G}}}'_{i}}
    -{\mathbf{F}}^\mathrm{H}\Delta{{\mathbf{{G}}}'_{j}}
     \right\|^2_\mathrm{F}\\  
    &\overset{(a)}{\leq} \left\| {\mathbf{F}}^\mathrm{H}{{\mathbf{{G}}}'_{i}} \right\|^2_\mathrm{F} 
    +\left\| {\mathbf{F}}^\mathrm{H}{{\mathbf{{G}}}'_{j}} \right\|^2_\mathrm{F} \\
    &\overset{(b)}{\leq}  \sum_{n\in \mathcal{N}} \left\|\mathbf{f}_n\right\|^2_2 \cdot
    \left\|\Delta \mathbf{G}_{i,n}\right\|^2_\mathrm{F} 
    +\sum_{n\in \mathcal{N}} \left\|\mathbf{f}_n\right\|^2_2\cdot
    \left\|\Delta \mathbf{G}_{j,n}\right\|^2_\mathrm{F} \\
    &=\sum_{n\in \mathcal{N}}p_n({\epsilon}_{i,n}^2+{\epsilon}_{j,n}^2)
    \triangleq \hat{\epsilon}^2_{ij},
  \end{aligned}
\end{equation}
where $p_n$ refers to the power allocation for the $n$-th subcarrier,
satisfying $\sum_{n\in\mathcal{N}}p_n=1$.
Both inequalities $(a)$ and $(b)$ are valid according to the triangle inequalities of 
the Frobenius norm.
Accordingly, 
we can convert the infinitely many constraints into equivalent 
forms with only a finite number of constraints.
First, we reformulate \eqref{eq:huge_constraint1} into quadratic form as
\begin{equation}
  \begin{aligned}
    \eqref{eq:huge_constraint1} = & \ \hat{\mathbf{D}}_{ij}\boldsymbol{\Theta}\hat{\mathbf{D}}_{ij}^\mathrm{H}
    +2\Delta{\mathbf{D}}_{ij}\boldsymbol{\Theta}\hat{\mathbf{D}}_{ij}^\mathrm{H}\\
    &+\Delta{\mathbf{D}}_{ij}\boldsymbol{\Theta}\Delta{\mathbf{D}}_{ij}^\mathrm{H}
    - d\mathbf{I}_{N_p}\succeq \mathbf{0}, \forall \Delta\mathbf{D}_{ij}\in\boldsymbol{\varOmega_{ij}},
  \end{aligned}
\end{equation}
where $\boldsymbol{\Theta}\triangleq {\boldsymbol{\theta}'}
{\boldsymbol{\theta}'}^\mathrm{H}$ is imparted 
as a substitution for the quadratic term.
The equivalence holds on the condition of the following constraints:
\begin{align}
  \label{eq:LMIs_1}
  |\theta_m| = 1 \stackrel{\boldsymbol{\Theta}\triangleq {\boldsymbol{\theta}'}
  {\boldsymbol{\theta}'}^\mathrm{H}}{\Longleftrightarrow } 
  \left\{\begin{array}{*{1}{l}}
    \operatorname{Diag}(\boldsymbol{\Theta}) = \boldsymbol{1}_{M+1}, \\
    \operatorname{rank}(\boldsymbol{\Theta}) = 1, \\
    \boldsymbol{\Theta} \succ \mathbf{0}.
  \end{array}\right.
\end{align}
The introduced rank-one constraint still leaves the problem NP-hard
and will be address later.
Now, to tackle the infinite number of quadratic matrix inequalities (QMIs), we resort to the  
S-procedure Lemma \cite[Proposition 3.4]{luo2004multivariate} 
to convert the constraints into a finite number of LMIs.
\begin{lemma}
  Given the bounded variable $\mathbf{X}$ with 
  $\mathrm{Tr}\left(\mathbf{E}\mathbf{X}\mathbf{X}^\mathrm{H}\right)\leq 1$,
  the following QMI 
  \begin{align}
    \mathbf{X}^\mathrm{H}\mathbf{A}\mathbf{X}+\mathbf{X}^\mathrm{H}\mathbf{B}
    +\mathbf{B}^\mathrm{H}\mathbf{X}+\mathbf{C}\succeq \mathbf{0}, \forall \mathbf{X}
  \end{align}
  holds if and only if there exists $ q\geq 0$ such that
  \begin{align}
    \label{S-proce}
   \quad
    \left[\begin{array}{cc}
      \mathbf{C} & \mathbf{B}^H \\
      \mathbf{B} & \mathbf{A}
      \end{array}\right]-q\left[\begin{array}{cc}
      \mathbf{I}_n & \mathbf{0} \\
      \mathbf{0} & -\mathbf{E}
      \end{array}\right] \succeq \mathbf{0},
  \end{align}
  where $\mathbf{A}, \mathbf{E} \in \mathbb{H}^{m\times m}, \mathbf{X}, 
  \mathbf{B} \in \mathbb{H}^{m \times n}$, and $\mathbf{C} \in \mathbb{H}^{n\times n}$.
\end{lemma}
\noindent By substituting \eqref{eq:LMIs_1} into \eqref{S-proce},
the constraint is transformed accordingly as
\begin{align}
  \label{exConB1}
  \mathbf{S}^\mathrm{(a)}_{ij}{\boldsymbol{\Theta}}
  {\mathbf{S}^\mathrm{(a)}_{ij}}^\mathrm{H}+
  \mathbf{Q}^\mathrm{(a)}_{ij}
  \succeq \mathbf{0},
\end{align}
where $\mathbf{S}^\mathrm{(a)}_{ij}\triangleq 
[\hat{\mathbf{D}}^\mathrm{H}_{ij}, \mathbf{I}_{M+1}]^\mathrm{H}$ and
$\mathbf{Q}^\mathrm{(a)}_{ij}= \operatorname{blkdiag}\left(-(q^\mathrm{(a)}_{ij}+ d)\mathbf{I}_{N_p},
{q^\mathrm{(a)}_{ij}}{\hat{\epsilon}^{-2}_{ij}}\mathbf{I}_{M+1}\right)$. Here $q^\mathrm{(a)}_{ij}\geq 0$
is an auxiliary variable.

The QoS constraint of \eqref{eq:SNRconstraint} is reformulated according to \eqref{eq:LMIs_1}:
\begin{align}
  \label{eq:QoScon}
  \min_{\Delta\mathbf{G}'}~\sum_{n\in\mathcal{N}}\log_2\left(1+
  \frac{\mathbf{f}_n^{\mathrm{H}}\mathbf{G}_{k,n}^\prime\boldsymbol{\Theta}
  {\mathbf{G}_{k,n}^{\prime\mathrm{H}}}\mathbf{f}_n}{\sigma_n^2}\right)\geq
  R_k, k\in \mathcal{K}.
\end{align}
Although \eqref{eq:QoScon} is convex with respect to $\boldsymbol{\Theta}$, 
it involves infinitely many log-quadratic constraints due to
the continuous channel uncertainty set.
To handle this challenge, we introduce an auxiliary optimization 
$\gamma_{k,n},k\in\mathcal{K},n\in\mathcal{N}$, and reformulate the QoS constraint as
\begin{numcases}{\eqref{eq:QoScon}\Leftrightarrow}
  \sum_{n\in\mathcal{N}}\log_2\left(1+
  \gamma_{k,n}\right)\geq \label{eq:qos1}
  R_k,
  \\
  \inf_{\Delta\mathbf{G}'}\{{\mathbf{f}_n^{\mathrm{H}}\mathbf{G}_{k,n}^\prime\boldsymbol{\Theta}
  {\mathbf{G}_{k,n}^{\prime\mathrm{H}}}\mathbf{f}_n}/{\sigma_n^2}\}
  \geq \gamma_{k,n},\label{eq:qos2}
\end{numcases}
where $\inf\{\cdot\}$ returns the infimum of the set.
The equivalence holds by exploiting the monotonicity of the logarithm function.
It can be observed that the second set of inequalities of \eqref{eq:qos2} is also in
the form of QMI and can be converted into a set of LMIs 
through the S-procedure.
Specifically, the equivalent constraints can be expressed as
\begin{align}
  \label{eq:qoos2}
  \eqref{eq:qos2}\Leftrightarrow
  \mathbf{S}^\mathrm{(b)}_{k,n}
  \boldsymbol{\Theta}
  {\mathbf{S}^\mathrm{(b)}_{k,n}}^\mathrm{H}+
  \mathbf{Q}^\mathrm{(b)}_{k,n}
  \succeq \mathbf{0},k\in\mathcal{K},n\in\mathcal{N},
\end{align}
where $\mathbf{S}^\mathrm{(b)}_{k,n}\triangleq 
[{\hat{\mathbf{G}}'_{k,n}}{}^\mathrm{H}\mathbf{f}_n, \mathbf{I}_{M+1}]^\mathrm{H}$ and
$\mathbf{Q}^\mathrm{(b)}_{k,n}= \operatorname{blkdiag}\left(-(q^\mathrm{(b)}
+{\gamma}_{k,n}\sigma^2_n),
{q^\mathrm{(b)}}{\epsilon}_{k,n}^{-2}\mathbf{I}_{M+1}\right)$,
while $q^\mathrm{(b)}\geq 0$ is also an auxiliary variable.

\subsection{Unit Modulus}
In this subsection, we aim to solve the non-convex rank-one 
constraint introduced in \eqref{eq:LMIs_1}.
This constraint is crucial for maintaining an equivalence of the transformation
but also leads to the NP-hard difficulty of the problem. 
To tackle this challenge, a common approach is to initially set aside the rank-one 
constraint and solve the problem as a convex optimization problem, i.e., 
semidefinite relaxation. 
Once an optimal solution is obtained, the rank-one solution is constructed 
using the Gaussian randomization method. However, this approach is 
computationally demanding and may not yield even a feasible solution to the 
original problem.
Instead, we propose a heuristic successive convex approximation (SCA)-based method to 
improve the obtained rank-one solution iteratively.
First, we rewrite the 
rank-one constraint in an equivalent form:
\begin{align}
  \label{equi-rank}
  \operatorname{rank}(\boldsymbol{\Theta}) =1 {\Leftrightarrow } 
  \left\| \boldsymbol{\Theta} \right\|_{2} \geq M,
\end{align}
where $\left\|\boldsymbol{\Theta}\right\|_{2}$ represents the spectral norm of 
$\boldsymbol{\Theta}$, which is defined as the maximum singular value of $\boldsymbol{\Theta}$. 
It is important to note that
the condition ${\operatorname{Diag}(\boldsymbol{\Theta}) = \boldsymbol{1}_M}$ is crucial 
to ensure the equivalence, and it can be found in \eqref{eq:LMIs_1}.
Specifically, we utilize the property of matrices that the sum of eigenvalues 
is equal to the trace of the matrix. Therefore, the right-hand side term in the transformation 
originates from $\operatorname{Tr}(\boldsymbol{\Theta}) = 
\sum_{i} \lambda_{i} \geq \max_{i}\left\{\lambda_{i}\right\}$, 
where equality holds if and only if $\boldsymbol{\Theta}$ has unit rank \cite{yu_robust_2020}.
Here, $\lambda_i$ is the $i$-th singular value of $\boldsymbol{\Theta}$. 
By introducing a lower bound on the spectral norm, 
we can enforce the rank-one constraint. However, 
the spectral norm is a convex function which makes the 
inequality constraint non-convex. To handle it, 
we first establish a lower bound of the spectral norm 
adopting its first-order Taylor expansion as follows:
\begin{equation}
  \begin{aligned}
    \|\boldsymbol{\Theta}\|_2 \geq &\left\|\boldsymbol{\Theta}^{(t)}\right\|_2\\
    &+\operatorname{Tr}\left[{\boldsymbol{v}^{(t)}}
    {\boldsymbol{v}^{(t)}}^{\mathrm{H}}\left(\boldsymbol{\Theta}-
    \boldsymbol{\Theta}^{(t)}\right)\right]
     \operatorname{Tr}\left({\boldsymbol{v}^{(t)}}
     {\boldsymbol{v}^{(t)}}^{\mathrm{H}}
     \boldsymbol{\Theta}\right),
    \end{aligned}
\end{equation}
where ${\boldsymbol{v}^{(t)}}$ is the eigenvector in
the $t$-th iteration of SCA corresponding to the maximum eigenvalue of
$\boldsymbol{\Theta}^{(t)}$ from the last iteration,
we can obtain a convex approximation constraint as follows 
\begin{align}
  \label{eq:spectral_constraint}
  \operatorname{Tr}\left(\boldsymbol{v}^{(t)}
  {\boldsymbol{v}^{(t)}}^{\mathrm{H}}
  \boldsymbol{\Theta}\right) \geq M.
\end{align}
As such, a convex approximation of the non-convex rank-one 
constraint is constructed. 

Then, we employ a penalty-based method inspired by \cite{wright1999numerical, 9013643} 
to enforce the transformed rank-one constraint. Specifically, we augment a 
penalty term to the objective function as
\begin{align}
  f(d, \boldsymbol{\Theta}) = -d + \frac{1}{\varrho} \left(\operatorname{Tr}\left(\boldsymbol{v}^{(t)} 
  {\boldsymbol{v}^{(t)}}^{\mathrm{H}}\boldsymbol{\Theta}\right) - M \right),
\end{align}
where $\varrho \rightarrow 0^{+}$ is a penalty factor that penalizes the violation of the rank-one constraint.
By adopting the aforementioned transformation and approximation, 
the original optimization problem is then converted into
\begin{IEEEeqnarray}{lcll}
  \mathcal{P}_2\text{:}~ &  \underset{d, {\boldsymbol{\Theta}},\gamma_{k,n}}{\min }   ~                                               & f(d, \boldsymbol{\Theta})   & 
  \label{eq:penalty_problem}  \\
  &\text{s.t.}                                                                       & \operatorname{Diag}(\boldsymbol{\Theta})=\boldsymbol{1}_M;\boldsymbol{\Theta} \succeq \mathbf{0},
                      \IEEEyessubnumber\label{eq:newcons}\\
                      &                                                             & \eqref{exConB1},\eqref{eq:qos1},\eqref{eq:qoos2}.
                    \IEEEyessubnumber
\end{IEEEeqnarray}
The optimization problem $\mathcal{P}_2$ is
converted into a standard convex
semi-definite program (SDP) and hence can be optimally
solved by existing convex optimization programming solvers such as CVX \cite{grant2014cvx}.
The overall algorithm proposed in this section is summarized in 
\textbf{Algorithm~1}. 
By iteratively adjusting the penalty factor and solving 
the converted optimization problem $\mathcal{P}_2$,
we can monotonically decrease the value of the objective.
In this way, the matrix of the coefficients of RIS converges to
a stationary point to $\mathcal{P}_2$ in polynomial time \cite{7547360}.
Furthermore, the computational complexity of each 
iteration of the proposed algorithm is given by
$\mathcal{O}\left(\log{\frac{1}{\varepsilon}}\sqrt{M+1} \left(
  K^2N^2M^2+K^3N^3
\right)\right)$ \cite[Th. 3.12]{122},
where $\mathcal{O}$ is the big-O notation.





\begin{algorithm}[tb] 
  \caption{Map Construct Algorithm}
	\label{alg:algorithm1}
  \hspace*{0.02in} {\bf Input:} 
  channel matrices $\mathbf{G}_k$ with errors,
  for all $i\in\mathcal{I}$;
   initiate $\boldsymbol{\Theta}^{(0)}$ with random phase. \\
  \hspace*{0.02in} {\bf Output:}
  output optimal $\boldsymbol{\theta}_o$
	\label{alg:1}
	\begin{algorithmic}[1]
    \STATE Set iteration index $r = 0$;
    \STATE Calculate $\mathbf{D}_{ij}$ by channel matrices;
    \REPEAT
    \STATE Solve problem $\mathcal{P}_2$ for given $\boldsymbol{\Theta}^{(r)}$,
      to update $\boldsymbol{\Theta}^{(r+1)}$
    \STATE Given $r+1$ for $\boldsymbol{\Theta}^{(r+1)}$, obtain the eigenvector $\boldsymbol{v}^{(t+1)}$
      corresponding to the largest eigenvalue;
    \STATE Update the penalty factor $\varrho$;
    \STATE Update $r\leftarrow r + 1$;
    \UNTIL{$|d^{(r+1)} - d^{(r)}|\leq \varepsilon$};
    obtain the optimal $\boldsymbol{\Theta}_o$
    \STATE Decompose $\boldsymbol{\Theta}_o=\boldsymbol{\tilde{\theta}}_o
    \boldsymbol{\tilde{\theta}}_o^\mathrm{H}$, obtain 
    $\boldsymbol{{\theta}}_o = \boldsymbol{\tilde{\theta}}_{o,\mathcal{M}}/\boldsymbol{\tilde{\theta}}_o(\rm{end})$.
  \end{algorithmic} 
\end{algorithm}

\section{CRLB Analysis}
In the last section, we qualitatively highlight the characteristic called ``correlation dispersion''.
In this section, we focus on a mathematical interpretation of this characteristic 
through statistical analysis. Our objective is to derive
the CRLB for 
the proposed positioning framework.

We omit the block division in Section II 
to explore the continuous bound of positioning 
performance.
In our work, we incorporate some prior information to achieve effective positioning,
wherein localization is essentially realized 
based on two sources of positioning information.
For a user located at $\mathbf{p}_0 = [x_0,y_0,z_0]$, the prior knowledge
can be readily obtained through prediction or rough positioning methods and be
modelled as Gaussian distribution \cite{8739375}:
\begin{align}
  \label{eq:pri_info}
  f(\mathbf{p}) = \frac{1}{(2\pi)^\frac{3}{2}\det^{\frac{1}{2}}(\boldsymbol{\Sigma}_0)}
  \exp\left(-\frac{1}{2}(\mathbf{p}-\mathbf{p}_0)^\mathrm{T}
  \boldsymbol{\Sigma}_0^{-1}(\mathbf{p}-\mathbf{p}_0)\right),
\end{align}
where $\boldsymbol{\Sigma}_0 \in \mathbb{R}^{3\times 3}$ represents the covariance matrix
(usually large in terms of positive definiteness).
On the other hand, the information provided by the CFR positioning method can be   
expressed as circularly symmetric complex Gaussian distribution:
\begin{align}
  \label{eq:cfr_info}
  f(\mathbf{r}) = \mathcal{CN}_{\boldsymbol{\Gamma}_0}(\mathbf{r}-\mathbf{r}_0),
\end{align}
where $\mathcal{CN}(\mathbf{x})=\frac{1}{\pi^{N_p}\det(\mathbf{A})}
\exp\left(-\mathbf{x}^\mathrm{H}
\mathbf{A}^{-1}\mathbf{x}\right)$ is the probability density function (pdf) of complex Gaussian distribution with 
covariance matrix $\mathbf{A}$.
The two distributions describe distinct 
aspects for the user at location $\mathbf{p}_0$ and 
need to be harmonized in the same domain for further analysis.
To reveal the advantage of integrating RIS, 
we opt to convert the prior knowledge \eqref{eq:pri_info} 
into the CFR domain. 
This transformation is systematically executed through a 
mapping function $\mathbf{r} = \mu(\mathbf{p})$ that establishes relationship between 
physical locations and positioning parameters, often involving angles and distances. 
However, with the introduction of RIS, exact expression can
be intricate due to multiple links
and may vary with different reflection coefficients of the RIS.

As illustrated in Fig.~\ref{fig3}, there is no straightforward connection between 
the physical space and the positioning space.
Even with a known closed-form expression for $\mu(\cdot)$,
evaluating $g(\mathbf{r}_0) = f(\mu^{-1}(\mathbf{r}_0))$ can be 
extremely computationally demanding. 
To address this challenge, we employ 
the kernel density estimator technique \cite{1190751} to estimate
the distribution of the prior information in the CFR domain.
Specifically, the commonly adopted 
kernel density estimation assumes 
that the probability density is a smoothed version of the empirical samples.
For a collection of $M$ measured data samples $\{\mathbf{r}_i\}_{i=1}^M$,
the estimate $\hat{g}(\mathbf{r})$ of the underlying pdf ${g}(\mathbf{r})$ is 
the average of radial kernel functions centered on the $L+1$ measured data samples
\begin{align}
  \label{eq:kernel}
  \hat{g}(\mathbf{r}) = \frac{1}{L+1}\sum_{i=0}^{L}K(\mathbf{r}-\mathbf{r}_i).
\end{align}
Here, we assume $K(\cdot)$ to be a complex-valued Gaussian kernel:
\begin{align}
  K(\mathbf{r}) = \mathcal{CN}_{\boldsymbol{\Sigma}}(\mathbf{r}),
\end{align}
and its covariance matrix (or kernel width) \cite{1190751}, is defined as $\boldsymbol{\Sigma}$.
As $L\rightarrow \infty$, it can be demonstrated that $\hat{g}(\mathbf{r})$ converges to the true density
under certain conditions~\cite{parzen1962estimation}.
For the empirical samples in \eqref{eq:kernel}, we choose $L$ locations that are physically
close to the expected location $\mathbf{r}_0$. For example, if $N_1$ in Fig.~\ref{fig4} seeks
localization through GNSS, $N_2\sim N_4$ are considered the relevant samples.

We adopt the log-linear pooling \cite{genest1984characterization} to aggregates the 
two opinions using a geometric average:
\begin{align}
  h(\mathbf{r}) = \frac{f(\mathbf{r})\cdot \hat{g}(\mathbf{r})}
  {\int_{\mathbb{C}}{f(\mathbf{r})\cdot \hat{g}(\mathbf{r})}\mathrm{d}\mathbf{r}}.
\end{align}
The form of $h(\mathbf{r})$ can be derived with the following lemma.
\begin{lemma}
  The pdf of the renewed distribution is given by $(L+1)$-term Gaussian mixture
  \begin{align}
    \label{eq:intPdf}
    h(\mathbf{r}) = \sum_{i=0}^{L} 
    {w_i}\cdot\mathcal{CN}_{\mathbf{E}}(\mathbf{r}-{\mathbf{e}}_i),
  \end{align}
  where the weight $w_i$,
  the positive definite matrix $\mathbf{E}$, and the $N_p\times 1$ vector $\mathbf{e}_i$
  is defined as 
  \begin{align}
    w_i =& \ \frac{\mathcal{CN}_{\mathbf{C}}
    (\mathbf{r}_0-\mathbf{r}_i)}
  {\sum_{j=0}^{L}\mathcal{CN}_{\mathbf{C}}
  (\mathbf{r}_0-\mathbf{r}_j)},\\
  \label{eq:mixCov}
  \mathbf{C} =& \ \boldsymbol{\Sigma}+\boldsymbol{\Gamma}_0,
  \mathbf{E} = \boldsymbol{\Gamma}_0(\boldsymbol{\Gamma}_0+
  \boldsymbol{\Sigma})^{-1}\boldsymbol{\Sigma},
  \mathbf{E}^{-1}\mathbf{e}_i = 
  \boldsymbol{\Gamma}_0^{-1}\mathbf{r}_0 + \boldsymbol{\Sigma}^{-1}\mathbf{r}_i.
  \end{align}
  \begin{proof}
    The key of Lemma 2 is the proof of the following identify
    \begin{align}
      \mathcal{CN}_{\boldsymbol{\Sigma}}(\mathbf{r}-\mathbf{r}_i)
      \cdot \mathcal{CN}_{\boldsymbol{\Gamma}_0}(\mathbf{r}-\mathbf{r}_0)
      =\mathcal{CN}_{\mathbf{E}}(\mathbf{r}-\mathbf{e}_i)
      \cdot \mathcal{CN}_{\mathbf{C}}(\mathbf{r}_0-\mathbf{e}_i),
    \end{align}
    for all $\mathbf{r}$ and $\mathbf{r}_i$.
    Please refer to \cite[Appendix D]{9100524} for a detailed proof.
  \end{proof}
\end{lemma}

To assess the performance of the proposed positioning method, 
the CRLB is derived based on the pdf \eqref{eq:intPdf}.
For an unbiased estimator, the estimation variance 
of the exact CFR $\mathbf{r}_0$ of location $\mathbf{p}_0$ is bounded by its CRLB, 
which is the inverse of the Fisher Information Matrix (FIM) $\mathbf{I}(\mathbf{r}_0)$.
Under regularity conditions, the FIM is defined by \cite{kay1993fundamentals}:
\begin{align}
  \label{eq:FIM}
  \mathbf{I}(\mathbf{r}_0) = -\mathbb{E}
  \left[\frac{\partial^2 \ln h(\mathbf{r})}{\partial \mathbf{r}_{0}
  \partial \mathbf{r}_{0}^\mathrm{T}}\right].
\end{align}
The second order partial derivative of \eqref{eq:intPdf} with respect to $\mathbf{r}_0$ 
is expressed as \eqref{eq:hessien} (in the bottom of next page).
\begin{figure*}[!b]
  \normalsize
  \setcounter{MYtempeqncnt}{\value{equation}}
  \hrulefill

  \setcounter{equation}{36}
  \begin{equation}
    \begin{aligned}
      \label{eq:hessien}\\
      \frac{\partial^2 \ln h(\mathbf{r})}{\partial \mathbf{r}_{0}\partial \mathbf{r}_{0}^\mathrm{T}} 
      =& 
      \frac{-\partial^2 \ln \left( \sum_{j=0}^{L}\mathcal{CN}_{\mathbf{C}}
      (\mathbf{r}_0-\mathbf{r}_j)\right)
      +{\partial^2 \ln \left(\mathcal{CN}_{\boldsymbol{\Gamma}_0}(\mathbf{r}-\mathbf{r}_0) \right)}
      +{\partial^2 \ln \left(\sum_{j=0}^{L}\mathcal{CN}_{\boldsymbol{\Sigma}}
      (\mathbf{r}-\mathbf{r}_j) \right)}}
      {\partial \mathbf{r}_{0}\partial \mathbf{r}_{0}^\mathrm{T}}
      \\
      =&-\frac{4\sum_{j=1}^{L}\mathcal{CN}_{\mathbf{C}}
      (\mathbf{r}_0-\mathbf{r}_j)
      {\mathbf{C}}^{-1}  
      (\mathbf{r}_0-\mathbf{r}_j)(\mathbf{r}_0-\mathbf{r}_j)^\mathrm{H}
      {\mathbf{C}}^{-1} 
      -2{\sum_{j=1}^{L}\mathcal{CN}_{\mathbf{C}}
      (\mathbf{r}_0-\mathbf{r}_j)}\mathbf{C}^{-1}
      }
      {\sum_{j=0}^{L}\mathcal{CN}_{\mathbf{C}}
      (\mathbf{r}_0-\mathbf{r}_j)} \\
      & + 
        \frac{4\left(\sum_{j=1}^{L}\mathcal{CN}_{\mathbf{C}}
        (\mathbf{r}_0-\mathbf{r}_j)
        {\mathbf{C}}^{-1}  
        (\mathbf{r}_0-\mathbf{r}_j)\right)
        \left(\sum_{j=1}^{L}\mathcal{CN}_{\mathbf{C}}
        (\mathbf{r}_0-\mathbf{r}_j)
        {\mathbf{C}}^{-1}  
        (\mathbf{r}_0-\mathbf{r}_j)\right)^\mathrm{H}
        }
        {\left(\sum_{j=0}^{L}\mathcal{CN}_{\mathbf{C}}
        (\mathbf{r}_0-\mathbf{r}_j)\right)^2}\\
        &-2\boldsymbol{\Gamma}_0^{-1} 
        +\frac{4\mathcal{CN}_{\boldsymbol{\Sigma}}
        (\mathbf{r}-\mathbf{r}_0)
        {\boldsymbol{\Sigma}}^{-1}  
        (\mathbf{r}-\mathbf{r}_0)(\mathbf{r}-\mathbf{r}_0)^\mathrm{H}
        \boldsymbol{\Sigma}^{-1} 
        -2\mathcal{CN}_{\boldsymbol{\Sigma}}(\mathbf{r}-\mathbf{r}_0)\boldsymbol{\Sigma}^{-1}
        }
        {\sum_{j=0}^{L}\mathcal{CN}_{\boldsymbol{\Sigma}}
        (\mathbf{r}-\mathbf{r}_j)} \\
        &-\frac{4\mathcal{CN}^2_{\boldsymbol{\Sigma}}
        (\mathbf{r}-\mathbf{r}_0)
        {\boldsymbol{\Sigma}}^{-1}  
        (\mathbf{r}-\mathbf{r}_0)
        (\mathbf{r}-\mathbf{r}_0)^\mathrm{H}{\boldsymbol{\Sigma}}^{-1}
        }
        {\left(\sum_{j=0}^{L}\mathcal{CN}_{\boldsymbol{\Sigma}}
        (\mathbf{r}-\mathbf{r}_j)\right)^2}.
    \end{aligned}
    \end{equation}
  \setcounter{equation}{\value{MYtempeqncnt}}
  \vspace*{4pt}
      \end{figure*}
Note that the constant terms are neglected.
It is evident that only the last two terms 
on the right-hand side are functions of variable $\mathbf{r}$.
In this paper, we leave out the case where $\mathbf{r}_i = \mathbf{r}_0,\forall i$.
For clarity of representation,
we consolidate the remaining invariable terms (with respect to $\mathbf{r}$) of
\eqref{eq:hessien} into $i(\mathbf{r}_0)$.
Thus, the FIM can be formulated by \eqref{eq:sfaksnd}.
\begin{figure*}[!t]
  \normalsize
  \setcounter{MYtempeqncnt}{\value{equation}}
  \setcounter{equation}{37}
  \begin{equation}
    \begin{aligned}
      \label{eq:sfaksnd}
      \mathbf{I}(\mathbf{r}_0) =& -i(\mathbf{r}_0)  
      +\int_{\mathbb{C}} 
      \frac{4\mathcal{CN}^2_{\boldsymbol{\Sigma}}
      (\mathbf{r}-\mathbf{r}_0)
      {\boldsymbol{\Sigma}}^{-1}  
      (\mathbf{r}-\mathbf{r}_0)
      (\mathbf{r}-\mathbf{r}_0)^\mathrm{H}{\boldsymbol{\Sigma}}^{-1}
      \cdot 
      \mathcal{CN}_{\boldsymbol{\Gamma}_0}(\mathbf{r}-\mathbf{r}_0)
      }
      {\sum_{j=0}^{L}\mathcal{CN}_{\boldsymbol{\Sigma}}
      (\mathbf{r}-\mathbf{r}_j)
      \cdot\sum_{j=0}^{L}\mathcal{CN}_{\mathbf{C}}
      (\mathbf{r}_0-\mathbf{r}_j)
      }\mathrm{d}\mathbf{r}\\
      &-\int_{\mathbb{C}} \frac{\left(
      {4\mathcal{CN}_{\boldsymbol{\Sigma}}
      (\mathbf{r}-\mathbf{r}_0)
      {\boldsymbol{\Sigma}}^{-1}  
      (\mathbf{r}-\mathbf{r}_0)(\mathbf{r}-\mathbf{r}_0)^\mathrm{H}
      \boldsymbol{\Sigma}^{-1} 
      -2\mathcal{CN}_{\boldsymbol{\Sigma}}(\mathbf{r}-\mathbf{r}_0)\boldsymbol{\Sigma}^{-1}
      }\right)
      \mathcal{CN}_{\boldsymbol{\Gamma}_0}
      (\mathbf{r}-\mathbf{r}_0)}
      {\sum_{j=0}^{L}\mathcal{CN}_{\mathbf{C}}
      (\mathbf{r}_0-\mathbf{r}_j)}  
      \mathrm{d}\mathbf{r}\\
      =& -i(\mathbf{r}_0)
      - \frac{\mathcal{CN}_{\mathbf{C}}(\mathbf{0})}
      {\sum_{j=0}^{L}\mathcal{CN}_{\mathbf{C}}(\mathbf{r}_0-\mathbf{r}_j)}
      \left(4\boldsymbol{\Sigma}^{-1}
        \mathbf{E}
      \boldsymbol{\Sigma}^{-1}
      -2\boldsymbol{\Sigma}^{-1}\right)\\
      &+\int_{\mathbb{C}} 
      \frac{4\mathcal{CN}^2_{\boldsymbol{\Sigma}}
      (\mathbf{r}-\mathbf{r}_0)
      {\boldsymbol{\Sigma}}^{-1}  
      (\mathbf{r}-\mathbf{r}_0)
      (\mathbf{r}-\mathbf{r}_0)^\mathrm{H}{\boldsymbol{\Sigma}}^{-1}
      \cdot 
      \mathcal{CN}_{\boldsymbol{\Gamma}_0}(\mathbf{r}-\mathbf{r}_0)
      }
      {\sum_{j=0}^{L}\mathcal{CN}_{\boldsymbol{\Sigma}}
      (\mathbf{r}-\mathbf{r}_j)
      \cdot\sum_{j=0}^{L}\mathcal{CN}_{\mathbf{C}}
      (\mathbf{r}_0-\mathbf{r}_j)
      }\mathrm{d}\mathbf{r}.
    \end{aligned}
    \end{equation}
  \setcounter{equation}{\value{MYtempeqncnt}}
  \hrulefill
  \vspace*{4pt}
  \end{figure*}
Recalling the relationship $\mathbf{r}_j = \mathbf{D}_j \boldsymbol{\theta}'$, 
it is important to note that the FIM is essentially a function of 
the reflection coefficients $\boldsymbol{\theta}'$. 
This implies that the performance of localization varies  
under different configurations of RIS. 
This property, not previously considered in the literature, holds the potential 
for achieving improved positioning performance.

In this section, our goal is to derive the minimum CRLB. 
Instead of directly obtaining the CRLB,
we focus on its inverse relation with the FIM, aiming to enhance the FIM, 
as suggested in~\cite{rao1973linear}. 
It is important to note that both CRLB and FIM are discussed in the scope of 
positive definiteness of matrix.
The close relationship between the FIM and 
the selection of $\mathbf{r}_j$ becomes evident.
Based on the assumption that all Gaussian components share the same covariance 
matrix $\boldsymbol{\Sigma}$, a fundamental conclusion can be drawn: 
at the point of maximum FIM, all $\mathbf{r}_j$ 
converge to the same optimal solution (or set), denoted as $\mathbf{r}^{\star}$, 
i.e., $\mathbf{r}_j = \mathbf{r}^{\star},\forall j =1,\cdots, L$.
Consequently, we reformulate the FIM as \eqref{eq:realCRLB}.
\begin{figure*}[!t]
  \normalsize
  \setcounter{MYtempeqncnt}{\value{equation}}
  \setcounter{equation}{38}
  \begin{equation}
    \label{eq:realCRLB}
    \begin{aligned}
      \mathbf{I}(\mathbf{r}_0; \boldsymbol{\theta}') 
      =& \frac{
      -2{L\mathcal{CN}_{\mathbf{C}}
      (\mathbf{r}_0-\mathbf{r}_j)}\mathbf{C}^{-1}
      }
      {L\mathcal{CN}_{\mathbf{C}}
      (\mathbf{r}_0-\mathbf{r}_j)
      +\mathcal{CN}_{\mathbf{C}}
      (\mathbf{0})} 
      +\frac{4L\mathcal{CN}_{\mathbf{C}}
      (\mathbf{r}_0-\mathbf{r}_j)
      {\mathbf{C}}^{-1}  
      (\mathbf{r}_0-\mathbf{r}_j)
      (\mathbf{r}_0-\mathbf{r}_j)^\mathrm{H}{\mathbf{C}}^{-1}  
      }
      {\left(L\mathcal{CN}_{\mathbf{C}}
      (\mathbf{r}_0-\mathbf{r}_j)+\mathcal{CN}_{\mathbf{C}}
      (\mathbf{0})\right)^2}\\
      &+2\boldsymbol{\Gamma}_0^{-1} 
      - \frac{\mathcal{CN}_{\mathbf{C}}(\mathbf{0})}
    {L\mathcal{CN}_{\mathbf{C}}(\mathbf{r}_0-\mathbf{r}_j)+\mathcal{CN}_{\mathbf{C}}(\mathbf{0})}
    \left(4\boldsymbol{\Sigma}^{-1}
      \mathbf{E}
    \boldsymbol{\Sigma}^{-1}
    -2\boldsymbol{\Sigma}^{-1}\right)\\
    &+\frac{4\mathcal{CN}_{\mathbf{C}}
    (\mathbf{0}) {\boldsymbol{\Sigma}}^{-1}  }
    {\left(L\mathcal{CN}_{\mathbf{C}}(\mathbf{r}_0-\mathbf{r}_j)
    +\mathcal{CN}_{\mathbf{C}}(\mathbf{0})\right)}
    \int_{\mathbb{C}} 
    \frac{\mathcal{CN}_{\boldsymbol{\Sigma}}
    (\mathbf{r}-\mathbf{r}_0)
    \mathcal{CN}_{\mathbf{E}}
    (\mathbf{r}-\mathbf{r}_0)\cdot
    (\mathbf{r}-\mathbf{r}_0)
    (\mathbf{r}-\mathbf{r}_0)^\mathrm{H}
    }
    {L\mathcal{CN}_{\boldsymbol{\Sigma}}
    (\mathbf{r}-\mathbf{r}_j)+\mathcal{CN}_{\boldsymbol{\Sigma}}
    (\mathbf{r}-\mathbf{r}_0)
    }\mathrm{d}\mathbf{r}{\boldsymbol{\Sigma}}^{-1}.
    \end{aligned}  
  \end{equation}
  \setcounter{equation}{\value{MYtempeqncnt}}
  \hrulefill
  \vspace*{4pt}
  \end{figure*}
It can be observed that 
when $\mathbf{r}_j$ approaches $\mathbf{r}_0$, the FIM converges to
\setcounter{equation}{39}
\begin{equation}
  \begin{aligned}
    \mathbf{I}_0 =& \lim_{\mathbf{r}_j\rightarrow \mathbf{r}_0}
    \mathbf{I}(\mathbf{r}_0; \boldsymbol{\theta}')\\
    =& 2\boldsymbol{\Gamma}_0^{-1} 
    +\frac{2}{L+1}\boldsymbol{\Sigma}^{-1}
    -\frac{2L}{L+1}\mathbf{C}^{-1}
    -\frac{4L}{(L+1)^2}\boldsymbol{\Sigma}^{-1}
    \mathbf{E}
    \boldsymbol{\Sigma}^{-1}.
  \end{aligned}
\end{equation}
Additionally, as $\|\mathbf{C}^{-\frac{1}{2}}\left(\mathbf{r}_j -\mathbf{r}_0\right)\|_2$ 
tends to infinity, the FIM becomes:
\begin{align}
  \label{eq:rjinfty}
  \mathbf{I}_\infty =\lim_{\|\mathbf{C}^{-\frac{1}{2}}\left(\mathbf{r}_j 
  -\mathbf{r}_0\right)\|_2\rightarrow \infty}
  \mathbf{I}(\mathbf{r}_0; \boldsymbol{\theta}') = 
  2\boldsymbol{\Gamma}_0^{-1} 
  +{2}\boldsymbol{\Sigma}^{-1}.
\end{align}

Accounting for the definition of the limit operation,
we can deduce that both $\mathbf{r}_0$ and 
the set $\{\mathbf{r}_j:\mathbf{I}(\mathbf{r}_0; \boldsymbol{\theta}') = \mathbf{I}_\infty\}$ 
are stationary points of $\mathbf{I}(\mathbf{r}_0; \boldsymbol{\theta}')$.
Notably, the latter set 
represents a suboptimal solution for maximizing the FIM.
This conclusion aligns with the intuitive understanding 
that positioning performance improves when interference 
from other locations is minimized, and worsens when all 
locations are closely grouped together.
However, obtaining the global optimum can be challenging due to the complex expression involved. 
The conventional method of setting the partial derivative of 
$\mathbf{I}(\mathbf{r}_0; \boldsymbol{\theta}')$ with respect to 
$\mathbf{r}_j$ equal to~$\mathbf{0}$ leads to a transcendental equation, 
the solution of which is generally not in a closed form.
Moreover, considering the limitations in the capabilities of 
RIS, even if the optimal solution can be determined, 
it remains challenging to ascertain the corresponding 
reflection coefficients. Specifically, 
the following series of equations may result in an empty feasible set for $\boldsymbol{\theta}'$:
\begin{equation}
  \begin{aligned}
    \mathbf{D}_j\boldsymbol{\theta}' 
    = \mathbf{r}^{\star},
    \text{s.t.}~
    |\boldsymbol{\theta}^\prime_m| = 1,\forall j=1,2,\cdots, L ,\forall m\in\mathcal{M}.
  \end{aligned}
\end{equation}
Building upon the analysis, it is considered that separate neighboring locations
to improve positioning performance.
Subsequently, a CRLB for unbiased estimators is obtained as: 
\begin{align}
  \label{eq:crlbR}
  \mathrm{cov}(\hat{\mathbf{r}}_0)\geq \frac{1}{2}  
  \left(\boldsymbol{\Gamma}_0^{-1} 
  +\boldsymbol{\Sigma}^{-1}\right)^{-1}
  =\frac{1}{2} \mathbf{E},
\end{align}
where $\mathrm{cov}(\cdot)$ denotes the covariance matrix of the input vector.
For a geometric-based positioning method incorporated with the prior information, 
the CRLB is formulated as
\begin{equation}
  \begin{aligned}
    \label{eq:CRLBtra}
    \mathrm{cov}(\hat{\mathbf{p}}_0)\geq& 
    \left(\boldsymbol{\Gamma}^{-1} 
    +\boldsymbol{\Sigma}_0^{-1}\right)^{-1} \\
    = &
    \boldsymbol{\Gamma}
    \left(\boldsymbol{\Gamma} 
    +\boldsymbol{\Sigma}_0\right)^{-1}
    \boldsymbol{\Sigma}_0 
    =\boldsymbol{\Sigma}_0-
    \boldsymbol{\Sigma}_0
    \left(\boldsymbol{\Gamma} 
    +\boldsymbol{\Sigma}_0\right)^{-1}
    \boldsymbol{\Sigma}_0,
  \end{aligned}
\end{equation} 
where $\boldsymbol{\Sigma}_0
\left(\boldsymbol{\Gamma} 
+\boldsymbol{\Sigma}_0\right)^{-1}$ is also recognized as the famous Kalman gain.  
Here, $\boldsymbol{\Gamma}$
is the covariance matrix of the geometric positioning method, which measures 
the same spread as our method in the virtual positioning space.
Typically, the determinant of the covariance matrix is used 
to describe the spread of a random variable, known as generalized variance \cite{sengupta2004generalized}. 
In this context, the following relations hold:
\begin{align}
  \det(\boldsymbol{\Gamma}_0) &= \det(\boldsymbol{\Gamma}),\\
  \det(\boldsymbol{\Sigma}_0) &= \det\left(\boldsymbol{\Sigma}
  +\frac{1}{L+1}\sum_{i=0}^{L}(\mathbf{r}_j-\bar{\mathbf{r}})
  (\mathbf{r}_j-\bar{\mathbf{r}})^\mathrm{H}\right),
\end{align}
where $\bar{\mathbf{r}} = \sum_{i=0}^{L} \mathbf{r}_j/(L+1)$ represents the 
mean of the Gaussian mixture of \eqref{eq:kernel}.
The initial RIS coefficients are set the same for each element to 
simulate a usual NLOS link.
It is important to note that the CRLB in \eqref{eq:CRLBtra} 
represents the lowest bound achievable by conventional 
geometric-based methods. For specific methods like those relying on 
RSS or AoA detection \cite{AoA}, 
the corresponding Jacobian matrix of the transit function
must be multiplied by the FIM, a process that typically results 
in a degradation in performance.

\begin{figure}[!hbt]
  \centering
  \includegraphics[width=0.75\linewidth]{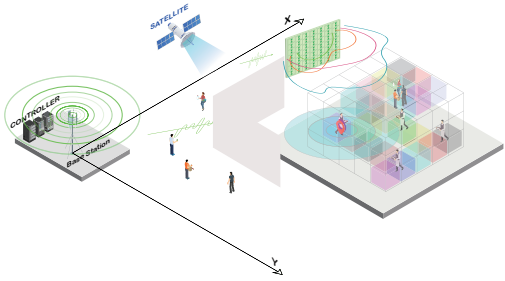}%
  \caption{Simulation environment setup.}
  \label{ststme} 
\end{figure}

\section{Simulations}

\subsection{Simulation Setup}
The considered simulation setup, as shown in Fig.~\ref{ststme},
involves a BS equipped with $N_T$ antennas at the center of the cell.
The precoding vector is calculated by 
the typical MRT algorithm to reduce computational complexity.
Randomly distributed single-antenna users (totaling $K$) are
placed in both outdoor and cuboid indoor areas,
with a higher indoor concentration. 
A RIS is appropriately deployed adjacent to the indoor area 
to establish transmission links for indoor users 
whose direct BS link is obstructed.
Additionally, GNSS provides coarse prior information for indoor users.
The cuboid indoor area is divided into blocks 
of dimensions $d \times d \times d$ for localization. 
In each realization of the channel, 
a specific subset of users with QoS requirements and another 
subset with positioning demands is randomly selected.

The cascaded channel matrix, $\mathbf{G}$, 
follows 
the IEEE 802.11b indoor propagation model \cite{1313233}. 
This model characterizes the channel as a frequency-selective fading channel, 
incorporating an exponentially decaying power delay 
profile with a root-mean-square delay spread. 
The initial channel of user-RIS-BS and user-BS are randomly generated 
at blocks where the physical distance between any two blocks exceeds 
the correlation distance. 
Subsequently, the channels for the remaining locations are 
obtained by interpolation \cite{interpolation} 
to preserve spatial consistency~\cite{zhu20193gpp}. 
The average number of taps in the impulse response is capped at $L_{\mathrm{max}}=6$. 
The covariance matrices for the prior 
information and additive noise in the frequency domain, 
as well as the covariance matrices for the prior information 
and equivalent noise in the physical space, 
are represented as
$\boldsymbol{\Sigma} = \sigma^2 \mathbf{I}_{N}$,
$\boldsymbol{\Gamma}_0 = \gamma_0^2 \mathbf{I}_{N}$,
$\boldsymbol{\Sigma}_0 = \sigma_0^2 \mathbf{I}_{3}$,
$\boldsymbol{\Gamma} = \gamma^2 \mathbf{I}_{3}$,
respectively.
The average received SNR for indoor users is defined as 
$\mathbb{E}_i[\left\|\mathbf{G}_i \boldsymbol{\theta}\right\|_2^2]/N\gamma^2$.
As the users are randomly distributed, 
we adopt the SNR of the user in the center of the cuboid area, 
as illustrated in Fig.~\ref{ststme}, 
as the reference SNR. This reference level, denoted as SNR, is 
utilized in the subsequent simulations.

For indoor users, 
we assume that the LoS link directly from the BS
is obstructed.
All the results are
obtained after averaging over $50$ independent random channel
realizations. 
The default system parameters are summarized in Table I.
Customized parameters related to each subsection will be specified subsequently.

\begin{table}[!thb]
  \label{tableI}
  \caption{System Parameters}
  \centering
  \begin{tabular}{|l|l|}
  \hline
  \multicolumn{1}{|c|}{Parameters} & \multicolumn{1}{c|}{Value}  \\
  \hline
      Carrier center frequency & $2.4$ GHz \\ \hline
      Number of sub-carriers and pilot carriers & $128$ and $8$ \\ \hline
      Default number of RIS elements & $32$ \\\hline
      Number of users, and with QoS requirements & $15$ dB and $4\sim 7 $ dB \\\hline
      Default QoS constraint, $R_k$     & $1.5$ bps/Hz  \\\hline
      Path loss exponent, $\alpha$ & $2.5$ \\\hline
      Bound of channel estimation error & $0.1$\\\hline
      Variance of $\sigma_0^2$ and $\gamma_0^2$  &  $5$ dB and $10$ dB  \\\hline
      Indoor region size ($\mathrm{m}^3$) & $9\times 9\times 3$ \\ \hline
      Penalty factor & $1\times 10^{-3}$\\ \hline
      Converge tolerance & $1\times 10^{-3}$\\\hline
  \end{tabular}
\end{table}


\subsection{Positioning Accuracy Evaluation}
\begin{figure}[!hbt]
  \centering
  \includegraphics[width=0.85\linewidth]{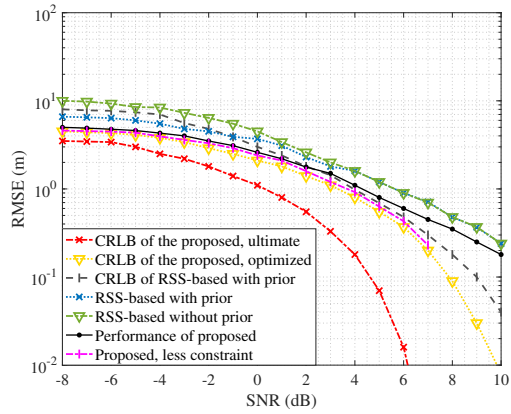}
  \caption{Performance evaluation and comparison of the proposed scheme,
  number of RIS elements $M=32$, number of pilot carriers $N_p=8$, and
  variance of the prior information $\sigma_0^2=5$ dB.}
  \label{performance}
\end{figure}

In Fig~\ref{performance}, 
we present and compare the positioning accuracy of different schemes in terms of 
root mean squared error (RMSE). 
The RSS-based positioning is realized with the methodology in \cite{zhang_rss_2021}.
The CRLB curves of both the conventional geometric-based method and 
the proposed two-stage scheme serve as baselines, 
establishing the bounds for positioning accuracy.
Notably, the CRLB curves of the proposed positioning 
framework outperform all other cases, especially in low SNR conditions.
This phenomenon showcases the superiority of the introduced 
correlation dispersion property.
The ultimate CRLB curve is obtained by \eqref{eq:crlbR}.
The number of Gaussian kernels in \eqref{eq:kernel} is set $L=12$.
The optimized CRLB curve is obtained by \eqref{eq:realCRLB},
which provides a more stringent bound.

In this subsection,
we set aside the QoS constraint in \eqref{eq:SNRconstraint}
to evaluate the performance of the embedded positioning function.
Moreover, we consider the case when
\eqref{eq:huge_constraint1} is reduced to
\begin{align}
  \eqref{eq:huge_constraint1}\Rightarrow \min_{\Delta\mathbf{G}'}~\left\|{{\mathbf{r}}_{i_0,\mathcal{P}} 
  -{\mathbf{r}}_{j,\mathcal{P}}}\right\|^2_2\geq d,
   j\in \mathcal{U}_{i_0}.
\end{align}
This single constraint means that we only consider the positioning of a specific
location.
This also leads support to our theoretical bound derivation of \eqref{eq:crlbR}.
For the case when all general constraints are considered, a degradation in positioning 
accuracy is clearly observed, especially when the SNR is relatively high.
In comparison to the conventional RSS-based positioning method, 
the proposed strategy shows superiority, particularly in scenarios with low SNR. 
Conversely, under high SNR conditions, geometric-based methods demonstrate 
improved positioning accuracy as LoS paths become more distinguishable.

In conclusion, the proposed framework performs 
better than conventional methods, 
capitalizing on distinct spatial 
consistencies in two different positioning spaces. 
In contrast, conventional geometric-based methods take advantage of 
comparatively less gain from the prior information, 
given the inherent similarity between the prior information and the methods themselves.
Building upon these observations, 
we have validated the introduced correlation dispersion characteristic.

\subsection{Positioning Performance Vesus QoS Requirements}
\begin{figure}[!hbt]
  \centering
   \includegraphics[width=0.85\linewidth]{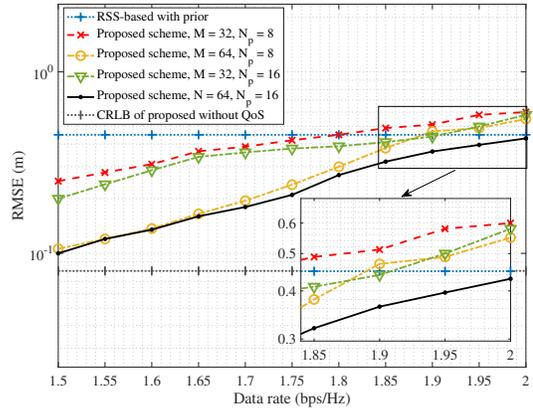}%
  \caption{Positioning accuracy versus communication requirements, variance of 
  prior information $\sigma_0^2=5$ dB, 
  SNR~=~$8$~dB.}
  \label{performance2} 
\end{figure}

Fig.~\ref{performance2} depicts the positioning performance in relation 
to the achievable data rate defined in~\eqref{eq:datarate}. 
Other system parameters follow the default settings and
the reference SNR is set to $10$ dB.
The number of communication users with specific requirements is randomly 
varied from $4$ to $7$ to simulate the dynamic of the communication system. 
As observed in Fig.~\ref{performance2}, there is a gradual degradation 
in positioning accuracy for the 
proposed method as QoS requirement becomes more stringent. 
This is attributed to both the RIS and the 
antenna array being optimized to enhance the received signal, 
resulting in higher SNR than initially designed. 
The embedded positioning function benefits 
from the improved propagation channel gain, 
leading to improved positioning accuracy. 
This observation emphasizes that the interconnection 
between the communication system and the embedded 
function cannot solely understood as a power allocation problem.
Owing to the dynamic number of QoS requirements,
the curves exhibit fluctuation with the
increase of required data rates, especially when 
the rates conditions are strict.

For the case with more RIS elements, i.e., $M=64$, $N_p = 8$, 
the embedded positioning function acquires an improvement
due to the redundancy in the RIS configuration for communication needs.
For the case with more pilot carriers, i.e., $M=32$, $N_p = 16$, 
the improvement is less pronounced  
as the overall SNR is low, and the capabilities of RIS are limited,
However, the additional dimensionality provides a more detailed resolution.
In the following, we will dig deeper into the relationship between 
the number of RIS elements and the number of pilot carriers.

\begin{figure}[!hbt]
  \centering
  \includegraphics[width=0.8\linewidth]{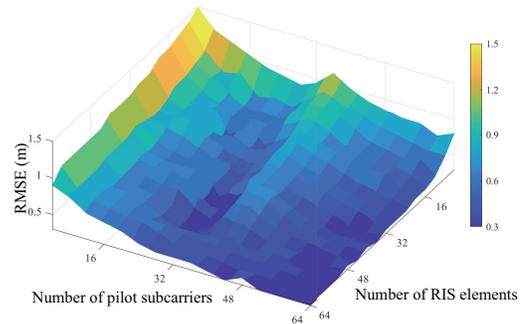}%
  \caption{Positioning accuracy versus the number of RIS elements and
  number of pilot carriers, SNR = $5$ dB, $\sigma_0^2 =  5$ dB, no QoS.}
  \label{performance3} 
\end{figure}

\subsection{Impact from Different System Settings}
In Fig. \ref{performance3}, we explore the impact of the number of
RIS elements and the length of the pilot subcarriers, 
without considering explicit QoS constraints.
The variance of the prior information is set to $5$ dB and the reference signal SNR is
set $5$ dB.
When the number of pilot carriers is fixed, 
a consistent positive impact on the positioning performance 
is observed with an increase in the number of 
RIS elements.
However, when the number of RIS elements is fixed, 
the monotonic relationship between the increase in RIS elements 
and the improvement in accuracy only holds when 
the number of RIS elements is considerably large.
Notably, there is a discernible set of 
stationary points where the accuracy of positioning 
deteriorates with the escalating number of pilot subcarriers.
Also, a gradual enhancement in positioning accuracy is 
evident with the increase of RIS elements.
This phenomenon is primarily attributed to the 
insufficiency of reshaping the distribution of 
location in the positioning space when the 
number of RIS elements is relatively small. 
Nevertheless, as the number of pilot subcarriers increases, 
a more detailed fingerprint can be obtained, 
thereby facilitating improved positioning accuracy.

In summary, optimal positioning performance is 
achieved at specific points where the number of RIS elements 
and the number of pilot subcarriers match optimally.
Beyond these points, 
there is a risk of degradation in performance, 
as they may counteract each other. Generally, 
an augmentation in the number of RIS elements 
consistently proves advantageous for the system.


\begin{figure}[!hbt]
  \centering
  \includegraphics[width=0.85\linewidth]{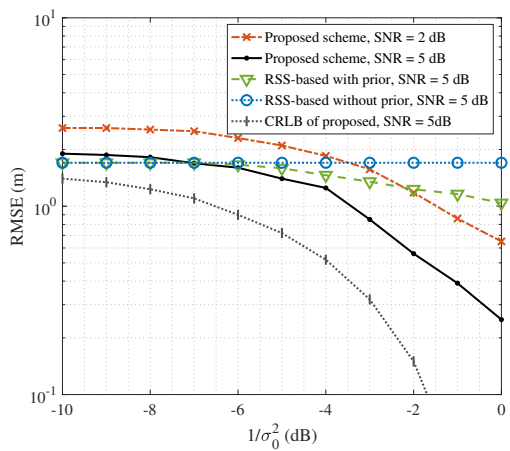}%
  \caption{Positioning accuracy versus the variance of prior 
  information, $\sigma_0^2 =  5$ dB, number of RIS elements $M=32$, 
  number of pilot carriers $N_p=8$.}
  \label{performance4} 
\end{figure}

\subsection{Positioning with Prior Information}
The combination of prior information in this paper 
results in an even smaller variance than both the 
variance of the prior information and the additive 
noise of the received signal. In Fig.~\ref{performance4},
we illustrate the positioning performance with 
decreasing variance of the prior information. 
When the variance of the prior information is relatively large, 
it contributes insufficient information to narrow down the 
potential area of the user or localize the user accurately. 
Consequently, the both curves of the proposed method 
reach to the upper bound that equals to 
the bound where there is no prior information. 
A noticeable improvement is observed earlier for 
the proposed scheme with SNR = $5$ dB compared to 
SNR = $2$ dB.
In contrast, the conventional RSS-based method achieves
similar improvement only when the prior 
information is relatively large. Specifically, 
to achieve meter-level accuracy,
a $3$ dB disparity is noted compared 
to the proposed scheme when SNR is set $5$~dB.

In conclusion, despite a noticeable gap due 
to the presence of numerous constraints,
the proposed methods consistently outperform the 
traditional approach when there is an increase
in the prior information.
It emphasizes the core idea of the proposed 
scheme that substantial improvement 
can be achieved when the location information 
captures different aspects.

\section{Conclusion}
In this paper, we explored the ability of RIS to reshape the 
wireless environment in the frequency domain.
Commencing with an integrated communication and positioning 
design based on CFR fingerprints, 
we proposed a two-stage positioning framework.
The introduction of the prior information and its transformation 
into the positioning domain unveils a 
novel spatial correlation dispersion property.
This property plays a crucial role in enhancing positioning performance, 
particularly in low SNR conditions. 
The CRLB of the proposed scheme was examined to theoretically validate this property.
Extensive simulation results demonstrated that the 
proposed schemes achieve significant performance in terms of
positioning accuracy, theoretical explanation, and communication assurance.

\appendices

\bibliographystyle{IEEEtran}
\bibliography{IEEEabrv, ref}


 




\vfill

\end{document}